\documentclass[10pt]{jfm}
\usepackage{epstopdf, epsfig}

\usepackage{amsmath,authblk,enumerate}
\usepackage{amssymb,amsbsy,comment}
\usepackage{soul}
\usepackage{subfig}
\usepackage{graphicx}
\usepackage{tikz,pgfplots}
\usepackage{mathrsfs}
\usepackage{bm}
\usepackage{xcolor}
\usepackage[hidelinks=true]{hyperref} 
\hypersetup{
    colorlinks   = true, %Colours links instead of ugly boxes
    urlcolor     = blue, %Colour for external hyperlinks
    linkcolor    = blue, %Colour of internal links
    citecolor   = red %Colour of citations
}

\def\d{\textrm{d}}
\def\Dm{\textrm{D}}
\def\dD{\partial \mathcal{D}}

\def\u{\bm{u}}

\def\F{\bm{F}}
\def\I{\bm{I}}

\def\k{\kappa}

\def\f12{\frac{1}{2}}

\def\q{\mathbf{\hat{q}}}

\def\u{\bm{u}}

\def\F{\bm{F}}
\def\I{\bm{I}}

\def\ga{\gamma}

\def\k{\kappa}
\def\g{\gamma}
\def\H{\mathscr{H}}

\def\O{\mathcal{O}}
\def\e{\textrm{e}}

\def\D{\mathcal{D}}
\def\sD{\mathscr{D}}
\def\dD{\partial\mathcal{D}}

\def\ekw{\e^{-i k x + \omega (k) t}}
\def\d{\textrm{d}}
\def\hq{\hat{\bm{q}}}
\def\hg{\tilde{\bm{g}}}

\def\hf{\tilde{\bm{f}}}

\def\C{\mathbb{C}}

\def\vg{\bm{g}}

\def\M{\mathbb{M}}

\def\I{\mathscr{I}}

\def\f{\bm{f}}

\def\v{\bm{v}}
\def\u{\bm{u}}
\def\y{\bm{y}}
\def\q{\bm{q}}
\def\g{\bm{g}}

\def\ez{\hat{e}_z}
\def\fot{\frac{1}{2}}

\def\His{\mathcal{H}}
\def\F{\mathcal{F}}
\def\tk{\tilde{k}}

\def\E{\mathscr{E}}
\def\Dt{\Delta t}
\def\tt{\tilde{t}}
\def\Lscr{\mathscr{L}}
\def\cC{\mathscr{C}}
\def\k{\varkappa}

\title{Accurate solution method for the Maxey-Riley equation, and the effects of Basset history}

\shorttitle{Accurate solution method for the MR equation}
\shortauthor{S. G. Prasath et. al.}

\author{S. Ganga Prasath \aff{1}
    , Vishal Vasan\aff{1} \corresp{\email{\texttt{vishal.vasan@icts.res.in}}} , Rama Govindarajan\aff{1}}
\affiliation{\aff{1}International Centre for Theoretical Sciences (ICTS-TIFR) Shivakote, Hesaraghatta Hobli, Bengaluru 560089, India.}

\date{}

\begin{document}
    \maketitle
    \begin{abstract}
        The Maxey-Riley equation has been extensively used by the fluid dynamics community to study the dynamics of small inertial particles in fluid flow. However, most often, the Basset history force in this equation is neglected. Analytical solutions have almost never been attempted because of the difficulty in handling an integro-differential equation of this type. Including the Basset force in numerical solutions of particulate flows involves storage requirements which rapidly increase in time. Thus the significance of the Basset history force in the dynamics has not been understood. In this paper, we show that the Maxey-Riley equation in its entirety can be exactly mapped as a forced, time-dependent Robin boundary condition of the one-dimensional diffusion equation, and solved using the Unified Transform Method. 

        We obtain the exact solution for a general homogeneous time-dependent flow field, and apply it to a range of physically relevant situations. In a particle coming to a halt in a quiescent environment, the Basset history force speeds up the decay as stretched-exponential at short time while slowing it down to a power-law relaxation, $\sim t^{-3/2}$, at long time. A particle settling under gravity is shown to relax even more slowly to its terminal velocity ($\sim t^{-1/2}$), whereas this relaxation would be expected to take place exponentially fast if the history term were to be neglected. An important mechanism for the growth of rain drops is by the gravitational settling of larger drops through an environment of smaller droplets, and repeatedly colliding and coalescing with them. Using our solution we estimate that the rate of growth rate of a rain drop can be gross overestimated when history effects are not accounted. We solve exactly for particle motion in a plane Couette flow and show that the location (and final velocity) to which a particle relaxes is different from that due to Stokes drag alone. 

        For a general flow, our approach makes possible a numerical scheme for arbitrary but smooth flows without increasing memory demands and with spectral accuracy. We use our numerical scheme to solve an example spatially varying flow of inertial particles in the vicinity of a point vortex. We show that the critical radius for caustics formation shrinks slightly due to history effects. 

        Our scheme opens up a method for future studies to include the Basset history term in their calculations to spectral accuracy, without astronomical storage costs. Moreover our results indicate that the Basset history can affect dynamics significantly.
    \end{abstract}

    \section{Introduction}
    Particles in fluid flows are ubiquitous~\citep{toschi2009}, examples include plankton in the ocean \citep{ardekani2016sedimentation,guseva2016history}, colloidal spheres in Stokes flow,  droplets in clouds~\citep{falkovich2002,croor2017,ravichandran2017}, suspended particulate matter in the atmosphere and various industrial flows. The simplest way to study these flows is to assume a one-way interaction, where the particle dynamics is dictated by the flow, but does not disturb the flow, nor is it influenced by other particles. This is a good assumption when particles are far smaller than all relevant flow length scales, and when the suspension of particles is dilute, as is the case in several of the examples given above. In this setting, the particle obeys the Maxey-Riley (MR)~\citep{maxey1983} equation, which is a force balance  in Lagrangian coordinates, given by
    \begin{align}
        \dot{\y} =& \ \v(t),\\
        R \dot{\v} =& \ \frac{\Dm \u }{\Dm t} - \frac{1}{S} \big(\v - \u \big) - \sqrt{\frac{3}{\pi S}} \bigg\{ \frac{1}{\sqrt{t}} (\v(0) - \u(0)) + \int_{0}^{t} \frac{( \dot{\v}(s) - \dot{\u}(s))}{{\sqrt{t-s}}} \d s \bigg\}, \label{eq:MRmain}\\
        \beta \equiv & \ \frac{\varrho_p}{\varrho_f},\ S \equiv \frac{1}{3} \frac{a^2/ \nu}{T}, \ R \equiv \frac{(1 + 2 \beta)}{3},
    \end{align}
    where $\y, \v$ are the vector position and velocity respectively of the particle, and $\u$ represents the (possibly spatially and temporally dependent) fluid velocity. We note here that when spatial variation in the fluid is not accounted for in (\ref{eq:MRmain}), the equation is known as the Basset-Boussinesq-Oseen equation~\citep{clift2005,parmar2018}. As our focus is towards general fluid flows we stick to the nomenclature of Maxey-Riley equations. Dots represent Lagrangian derivatives in time $t$ and $\Dm/\Dm t$ represents the material derivative with respect to the fluid velocity. Furthermore $\varrho_p, \varrho_f$ are the particle and fluid density respectively, $R$ the effective density ratio including added-mass effects, and $S$ the Stokes number, defined here as the ratio of particle relaxation time-scale to flow time-scale. The last term on the right  hand side of (\ref{eq:MRmain}) is known as the Basset history force or Basset history integral~\citep{basset1888treatise}. It is an integral force along the trajectory of the particle, from the initial time until time $t$, resulting from the differences in acceleration of particle and surrounding fluid, and will be of primary interest in this paper. An important contribution to the force balance was brought out by~\cite{farazmand2015} who showed the existence of a singular contribution from a non-zero relative velocity at the initial time-instant. Thus far in the literature, the implications of this singular term are largely unexplored. We will discuss its significance in later sections.

    \begin{figure}
        \centering
        \includegraphics[scale=0.8]{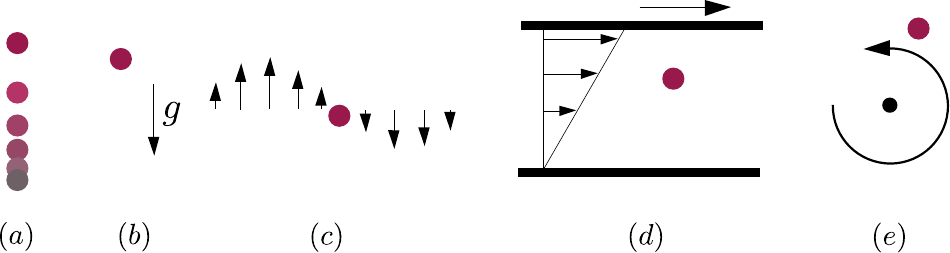}
        \caption{Situations for which analytical solutions including the Basset force are provided by the present approach. $(a)$ A relaxing particle, $(b)$ particle settling under gravity, $(c)$ particle in a oscillating field, $(d)$ particle in a shear flow, $(e)$ particle in a single point vortex.}
    \end{figure}

    Most studies concerning particle dynamics in the low Stokes number limit include contributions from the Stokes drag, the second term on the right-hand side in equation (\ref{eq:MRmain}), but the Basset history integral is often neglected. The coefficient of this term is $\O(S^{1/2})$ relative to the Stokes drag, so it is nominally negligible at small Stokes number, but the factor it multiplies could cause the effect to become important in physical flows. A major hurdle in evaluating the Basset history integral is the continually increasing memory cost associated in computing this term. The inclusion of the Basset history integral renders the MR equations to not represent a dynamical system, i.e., the future evolution of the particle motion depends not only on the current position and velocity, but also on the entire solution up to that time. Standard analytical techniques such as performing a Laplace transform are in general not useful in solving this system the full nonlinear equation. Moreover, even for the linear case, inverting the Laplace transform of a general function $G(t)$ with a kernel of the form $\sim1/\sqrt{t}$ multiplied to it does not lead to an explicit function of $t$. Thus researchers have resorted to quadrature schemes~\citep{van2011,daitche2013} or approximations to the history kernel~\citep{klinkenberg2014,elghannay2016development,parmar2018}. Since these techniques approximate the history-kernel and are not aimed at obtaining the true asymptotic behaviour, they neglect the aforementioned (most bothersome) singular contributions at $t=0$, which is a valid assumption when particles and fluid have the same initial velocity, which we emphasize, may not hold in many physical situations. By construction, the approximate schemes mentioned are polynomial order accurate. 

    In this paper we present $(i)$ analytical solutions, for the first time to our knowledge, of the complete Maxey-Riley equation, including the Basset history term, for particle dynamics in several canonical situations. The only solution we are aware of from earlier work is limited to a relaxing particle \citep{farazmand2015,langlois2015} where they use a Laplace transform approach. We show that our approach provides a simpler and a more general solution. $(ii)$ demonstrate a numerical approach, by means of an example, of particle dynamics near a point vortex, to solve the complete equations for any general flow to spectral accuracy, again for the first time to our knowledge. Our preference for spectral accuracy in the current work stems from our primary goal of investigating the relevance and impact of the Basset history integral on the dynamics of particles. Hence we seek to eliminate any possible source of error from lower-order schemes. 
 
    We do this by reformulating the equations of motion of the particle with the non-local history-dependence into a {\it local} problem for an extended dynamical system.  The main principle employed in deriving the extended dynamical system is a domain extension~\citep{vishal2018}. In other words, we represent the entirety of the MR equation as a boundary condition to the one-dimensional diffusion equation. The extended dynamical system couples three quantities that march in time: the position of the particle, the velocity of the particle and the field satisfying the diffusion equation (which effectively encodes the history term).  The reformulation enables us to write solutions explicitly for spatially uniform fluid flows. We derive explicit expressions for $(i)$ a particle relaxing in a stationary fluid. In this case, the equations simplify to a constant coefficient linear problem which can be explicitly solved using Laplace transform as well. However, even in this case, the solution by that approach is not an explicit function of time, but given as a convolution integral \citep{farazmand2015}; $(ii)$ a particle settling under gravity, $(iii)$ a particle in an oscillating background and $(iv)$ a particle in planar Couette flow. For short times, we find the particle velocity relaxes faster when the Basset history integral is accounted for than when it is not. Moreover, on accounting for the effect of the history integral, a cross-over time exists beyond which the particle relaxes with a power-law decay, whereas the decay would be exponential without.
    In the flows we study, we find that though the transient behaviour of particles in flows is different, the infinite time behaviour remains unchanged, consistent with the analysis of~\cite{langlois2015}. 

    The solution formula that we obtain is of a form where Watson's lemma~\citep{miller2006} may be applied and hence the large-time behaviour is readily computable. Our reformulation of the MR equation has the additional benefit of being local. This allows us to construct a novel numerical scheme that obviates the issue of rising memory storage. Since the effect of the Basset history integral is accounted for in terms of a dynamical variable, our method may also be employed in large simulations with restarts. In large numerical computations of turbulence, for example in cloud flows, the number of particles considered could be of the order of billions. Our method offers a way of including the Basset history integral into such simulations without an unmanageable increase in storage requirements.

    Close to a point vortex, upon neglecting the Basset history, the dynamics of particles has been shown to obey a boundary-layer structure \citep{ravichandran2015,deepu2017caustics}, where particles initially located within a critical radius $r_c$ are able to form caustics and evacuate the vicinity of the vortex rapidly, whereas particles initially outside this radius do not form caustics, and so their dynamics may be represented by a velocity field. We define caustics as space-time points where two particles can exist simultaneously with different velocities. We demonstrate our numerical method by studying how the Basset history integral affects caustic formation. We show that the history integral shrinks the critical caustic radius, $r_c$ for all Stokes numbers, but the scaling $r_c \sim \sqrt{S}$ is unchanged. Lastly, though all the boundary-bulk extension ideas are applied here to MR equations that describe dynamics of a spherical particle, we show that an exact map can be applied to particles with other geometries such as spheroids and disks in a flow where an analogous Basset history integral exists. We discuss the implications of our findings to turbulent scenarios, where the effect of the Basset history integral on particle clustering, is still not clearly understood.

    While our present study is devoted to the Maxey-Riley equations, we note that convective inertia effects become important beyond the diffusive time scale~\citep{lovalenti1993,lovalenti1993b,mei1992}, however small the Reynolds number. So the asymptotic behaviour we calculate below using the MR equation will only be valid up to this time scale, beyond which we must use the entire Navier-Stokes equation to get the correct behaviour. For example, in a particle relaxing to rest in quiescent flow, as calculated in the pioneering work of~\cite{lovalenti1993}, the relaxation behavour changes from a $1/\sqrt{t}$ decay, as predicted by the MR equation with Basset history, to a faster decay of $1/t^2$ at long times. In the present work, we assume a time-scale separation between the history effects arising out of unsteady inertial effects and the time-scale due to convective inertial corrections. As a consequence our long time asymptotic results are valid at time scales shorter than the convective scale $\nu/u_c^2$, where $u_c$ is a characteristic convective velocity. Two situations where there is at least two orders of magnitude of separation between the Stokes and convective time scales, and Basset history could be vital to understanding the relevant dynamics, are droplet growth in cloud flows, and sedimentation of marine snow. It is significant to note that that the technique developed here can be extended to account for the convective inertia as well, using a history kernel introduced by~\citep{lovalenti1993}, which will be addressed in future work.

    Before we dive into the details of the calculation, we describe our procedure to handle the Basset history term. It is known that the history integral is nothing but a half-derivative in time of the particle velocity relative to fluid motion and it is no surprise that the origins of this half-derivative has to do with the unsteady Stokes equation with particle momentum being diffused by viscosity. On the other hand, it is also known that the half-derivative is the Dirichlet-to-Neumann map for the diffusion equation on the half-line. Let us unpack the previous statement. Suppose $\g_0(t)$ is the Dirichlet condition for a function $ \q(x, t)$ that satisfies the diffusion equation for $x>0$ (with zero initial conditions). Then the half-derivative of $\g_0(t)$ is equal to the Neumann condition $\q_x(0,t)$. %This leads us to use the fact that half-derivative is nothing but the map between Dirichlet boundary condition and Neumann boundary condition to the diffusion or diffusion equation. 
    \noindent This fact allows us to relate the MR equation, which is an ordinary integro-differential equation, to a boundary condition to a diffusion equation.
    %We thus convert the MR equation, which is an ordinary integro-differential equation, into a boundary condition to the diffusion equation. 
    \noindent The boundary condition, which is the MR equation itself, is a time evolving Robin-type boundary condition and using the Unified Transform Method, as described below, we are immediately able to solve for the particle evolution for arbitrary time-dependent flows.

    \section{Diffusion equation on half-line}
    Before we discuss the MR equation itself, we first describe a few results pertaining to the $1$-D diffusion equation that are essential to our reformulation of the problem. The Unified Transform Method, alternatively known as the Fokas transform method, is employed in deriving the results in this section. These results are not new. Our intent is to introduce notation as well as the techniques that will be relevant in the following sections. Further details regarding the Unified Transform Method can be found in \citet{fokas2008unified}. For a brief introduction to the method see \citet{deconinck2014}.

    \begin{figure}
        \centering
        \includegraphics[scale=0.7]{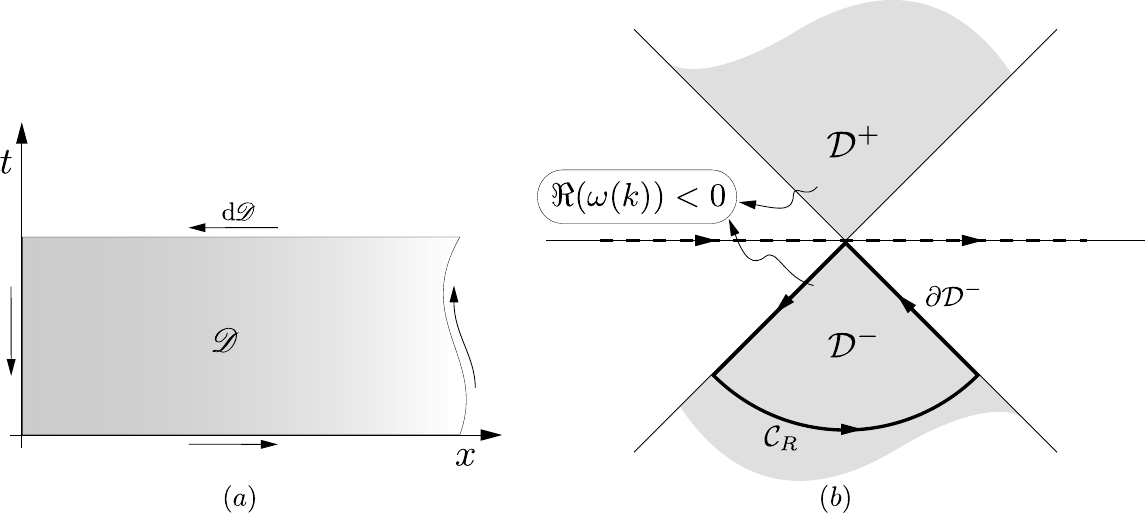}
        \caption{$(a)$ Region $\sD$ defined by $x \in [0,\infty), t \in [0,T]$ in which the diffusion equation evolves and $\d \sD$ is its boundary. $(b)$ Domain $\D^{\pm}$ where the integrating factor $\e^{\omega(k)t}$ decays. The boundary of this region in $\mathbb{C}^-$ is defined by contour $\dD^-$ and $\mathcal{C}_R$ whose various contributions help establish the relation between the Dirichlet and Neumann boundary conditions for the diffusion equation.}
        \label{fig:utm_cont}
    \end{figure}	

    Consider the 1-D diffusion equation on a half-line of `\textit{pseudo}'-space $x$ with a Dirichlet boundary condition given by $\g_o(t)$ at $x=0$ and let $\q(x,t)$ be the variable that gets diffused. Note that $x$ is a fictitious space, not to be confused with the physical space represented by $y$, so we call $x$ a pseudo-space. It is used only to establish the relationship between the Basset history integral and the Neumann boundary condition for the diffusion equation. This problem can be formulated in a time-interval $(0,T]$ as
    \begin{align}
        \q_t =& \ \q_{xx}, & \ x>0, t \in (0,T], \nonumber \\
        \q(x,0) =& \ 0, & \ x > 0, \label{eq:heat} \\
        \q(0,t) =& \ \g_o(t), & \ t \in [0,T]. \nonumber
    \end{align}
    Subscripts here denote partial derivatives and the evolution of the variable $\q(x,t)$ takes place in the $x-t$ plane in the domain $\sD$ shown schematically in figure~\ref{fig:utm_cont}$(a)$. As is typical for problems posed on a semi-infinite domain, we assume the field $\q(x,t)$ vanishes as $x\to\infty$ uniformly for all $t$. Though this problem may be solved using classical techniques, we employ the Unified Transform Method for reasons that will become evident in subsequent sections.

    We begin by rewriting the diffusion equation in its divergence form using the `\textit{local relation}'~\citep{deconinck2014,fokas2008unified,deconinck2017} 
    \begin{equation}
        (\ekw \q)_t - (\ekw (\q_x + ik \q))_x = 0, \ k \in \C, \label{eq:lr}
    \end{equation}
    which is valid in any small portion of $\sD$ where $\omega(k)=k^2$. The local relation is simply the diffusion equation multiplied by $\ekw$, as one can see by applying the product rule of differentiation to (\ref{eq:lr}).

    \subsection{The global relation and some notation}
    Next we integrate equation (\ref{eq:lr}), over the entirety of $\sD$ and employ the divergence theorem to obtain a `\textit{global relation}',
    \begin{align}
        \iint_{\sD} [ (\ekw \q)_t - (\ekw (\q_x + ik \q))_x ] \d x \ \d t =& 0,\\
        \Rightarrow \int_{\text{d}\sD} [\ekw \q \ \d x + \ekw(\q_x + ik \q)\ \d t ] =&\  0,\\
        \Rightarrow \int_{0}^{\infty} e^{-ikx} \q_o(x) \d x - \int_{0}^{\infty} \e^{-ikx+\omega(k) T } \q(x,T) \d x - \int_0^T \e^{\omega (k) t}(\q_x(0,t) + ik \g_o(t)) \d t =& \ 0.
    \end{align}
    The terms at the boundary of $\sD$ at infinity vanish since $\q(x,t)\to0$ as $x\to\infty$. Let us define the Fourier transform in pseudo-space as
    \begin{align}
        \hq_o (k) &= \int_0^\infty \e^{-ikx} \q_o(x) \ \d x, \\
        \hq (k,T) &= \int_0^\infty \e^{-ikx} \q(x,T) \ \d x,
    \end{align}
    and the time-transform of the boundary terms as
    \begin{align}
        \hg_o (\omega,T) &= \int_0^T \e^{\omega(k) t} \g_o (t) \ \d t,\label{eq:g0} \\
        \hg_1 (\omega,T) &= \int_0^T \e^{\omega(k) t} \q_x (0,t) \ \d t. \label{eq:g1}
    \end{align}
    Further we define $\D^{+}$ as \[ \D^{+} = \{ k\in \mathbb{C}\: :\: \Re(\omega(k))<0\: , \Im(k)>0\}, \]
    and $\D^{-}$ as \[ \D^{+} = \{ k\in \mathbb{C}\: :\: \Re(\omega(k))<0\: , \Im(k)<0\}, \]
    as shown in figure~\ref{fig:utm_cont}$(b)$. Note $\e^{\omega (k) t}$ is analytic, bounded and decaying for large $k$ in these domains, a property we shall use repeatedly. Finally we arrive at the global relation given by
    \begin{equation}
        \hq_o (k) - \e^{\omega(k) T} \hq (k,T) - \hg_1 - ik \hg_o = 0, \quad  \ k \in \C^-.
    \end{equation}
    Since the initial condition is zero everywhere on the real line, we have
    \begin{equation}
        \e^{\omega(k) T} \hq (k,T) + \hg_1 + ik \hg_o = 0, \quad  \ k \in \C^-.
        \label{eq:gr}
    \end{equation}

    \subsection{Dirichlet to Neumann map}
    Equation (\ref{eq:heat}) describes the Dirichlet boundary value problem for the diffusion equation. We now compute the associated Neumann condition, i.e., we ask what Neumann boundary condition would give the same solution as this Dirichlet condition in the entire domain. To do so we multiply equation (\ref{eq:gr}) by $ik \e^{-\omega(k)t}$ for $0<t<T$ and integrate over the contour $\dD^-$, the boundary of $\D^-$ shown in figure~\ref{fig:utm_cont}$(b)$.
    \begin{align}
        \int_{\dD^-} \bigg[ ik \int_0^T \e^{\omega (s-t)} \q_x (0,s) \ \d s - k^2 \int_0^T \e^{\omega (s-t) } \q (0,s) \ \d s + ik \e^{\omega(k) (T-t)} \q(k,T)  \bigg] \ \d k =& \ 0.
    \end{align}
    After the manipulations detailed in Appendix \ref{app:DNmap}, we obtain
    \begin{align}
        -\pi \q_x (0,t) - \sqrt{\frac{\pi}{t}} \q (0,0) - \int_{-\infty}^{\infty} \int_0^t \e^{-k^2 (t-s)} \dot{\q} (0,s) \ \d s \ \d k &= \ 0.
    \end{align}
    Setting $k^2 (t-s)=m^2$, where $m$ is real and evaluating the resulting Gaussian integral, we obtain the Dirichlet-Neumann map as
    \begin{align}	
        \q_x (0,t) =& - \sqrt{\frac{1}{\pi t}} \q (0,0) - \frac{1}{\sqrt{\pi}} \int_0^t \frac{\dot{\q} (0,s)}{\sqrt{t-s}} \ \d s.
        \label{DtoN}
    \end{align}
    The integral term above is the definition of the Riemann-Liouville half-derivative. This expression relating the Neumann condition for the diffusion equation and the Riemann-Liouville half-derivative forms the basis of our reformulation of the MR equations. More general connections between boundary-value problems to partial differential equations and fractional derivatives are explored in~\citet{vishal2018}.

    \section{Recasting the MR equation}\label{sec:MReqn}
    We may now use equation (\ref{DtoN}) to rewrite the Maxey-Riley equation (\ref{eq:MRmain}) in its entirety in a reference frame moving with the particle
    as
    \begin{align}
        \dot{\y} =& \ \q(0,t) + \u, \\
        \q_t(0,t) + \alpha \q(0,t) - \gamma \q_x(0,t) =& \ \f(\q(0,t),\y,t), \label{eq:MRFin} \\ \nonumber \\ 
        \alpha = \ \frac{1}{RS},\ \gamma = \frac{1}{R} \sqrt{\frac{3}{S}},\ \f(\q(0,t),\y(t),t) =& \ \bigg( \frac{1}{R} - 1 \bigg) \frac{\Dm \u }{\Dm t} - \q(0,t) \cdot \nabla \u, \label{eq:force}
    \end{align}
    where $\q(0,t) = \v(t) - \u(\y(t),t)$ is the relative velocity of the particle. In this choice of reference frame, with the history term on the left-hand side in (\ref{eq:MRFin}), we are left with a forcing function $\f$ which is local in time, whereas in the original form (\ref{eq:MRmain}) forcing appeared in a non-local manner. As $\q(0,t)$ and $\q_x(0,t)$ represent the Dirichlet and Neumann condition of a field satisfying the diffusion equation, we are naturally led to consider the following boundary-value problem
    \begin{figure}
        \centering
        \includegraphics[scale=0.9]{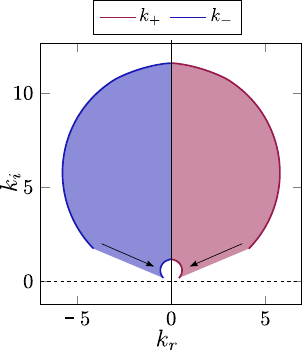}
        \caption{Region in $k \in \C^+$ where the poles $k_\pm$ from equation (\ref{eq:pole}) are located for Stokes number $S$ in the range $0.01-1$ and density ratios corresponding to $R$ in the range $1/3-5$. As the Stokes number increases the poles approach the real-line shown using arrows and similarly as we increase density ratio. Note that without  Basset history the poles would merely lie symmetrically on the $k_r$ line.}
        \label{fig:poles}
    \end{figure}
    \begin{align} \label{eq:HeatEqn}
        \q_t =& \ \q_{xx}, & \ x>0, t \in (0,T], \\
        \q(x,0) =& \ 0, & \ x > 0, \\
        \q_t(0,t) + \alpha \q(0,t) - \gamma \q_x(0,t) =& \ \f (\q(0,t),\y,t) , & \ t \in [0,T], \label{eq:BCmod} \\
        \dot{\y}(t) =& \ \q(0,t) + \u(\y(t)),& \ t\in[0,T],\label{eq:BCposition}\\
        \lim_{t \rightarrow 0} \q(0,t) =& \ \v_o,\\
        \y(0) =& \ \y_o. \label{eq:MRInit}
    \end{align}
    The MR equation (\ref{eq:MRmain}) thus manifests itself as a non-linear modified Robin boundary condition, on the time axis of figure~\ref{fig:utm_cont}$(a)$, to the diffusion equation. We do not yet know $\q(0,t)$ on this axis, i.e., the particle velocity in the relative frame of reference, for which we derive an expression in the next sub-section. From here on we work with this form of the MR equation and the global relation (\ref{eq:gr}) for the diffusion equation. Higher order corrections in particle size, the Fax\'en correction, etc. can also be accommodated in the forcing expression, $\f(\q(0,t),\y,t)$. Here $\u$ is a known velocity field. In the general case, one would couple the MR equation with a fluid model (such as the Navier-Stokes equation) to simultaneously resolve particle locations and fluid velocities. An example of this kind, in the form of particles around a point vortex, is studied in a later section.

    \subsection{Solution for time-dependent forcing}
    The form of the MR equation (\ref{eq:MRFin}-\ref{eq:force}), suggests the forcing function is both space and time-dependent. However it is instructive and useful to first study a case when the forcing is a function of time alone, i.e., $\f(t)$.  The ideas presented here will later be extended to more complicated scenarios in later sections.

    Using the definitions (\ref{eq:g0}-\ref{eq:g1}), we take the time-transform of the boundary condition  (\ref{eq:BCmod}) and after integrating by parts we get
    \begin{align}
        ( \e^{\omega(k) T} \q(0,T) - \v_o) + (\alpha - k^2) \hg_o - \gamma \hg_1 =& \ \hf.\label{eq:bcMR}
    \end{align}
    Here $\hf$ is the time-transform of the function $\f$. Using the global relation (\ref{eq:gr}) we eliminate $\hg_1$ from the above expression and multiply the resulting equality by $ik \e^{-\omega(k) t}$ for $0<t<T$ and finally integrate over $\dD^-$. This leads to
    \begin{align}
        \int_{\dD^-} ik \hg_o(\omega,T) \e^{-\omega(k) t} \ \d k =& \ \int_{\dD^-} ik \e^{-\omega(k) t} \bigg[ \frac{\hf - \e^{\omega T} \q(0,T) + \v_o - \gamma \e^{\omega T} \hq(k,T) }{(\alpha - k^2 + ik \gamma)} \bigg] \ \d k, \\
        \int_{\dD^-} ik \e^{-k^2 t} \int_0^T \e^{k^2 s} \q(0,s) \ \d s \ \d k =& \  \int_{\dD^-} \frac{ik \e^{-k^2 t} \big( \hf(k,T) + \v_o \big)} {(\alpha - k^2 + ik \gamma)} \ \d k \\ \nonumber
        & - \int_{\dD^-} ik \frac{ \e^{k^2 (T-t)} \big( \q(0,T) + \gamma \hq(k,T) \big)}{(\alpha - k^2 + ik \gamma)}  \ \d k,
    \end{align}
    where (in the second line) we have used the definition of $\hg_0$ and the fact that $\omega(k)=k^2$. We substitute $k^2 = il$ in the left-hand-side term above, which describes the boundary $\dD^-$, see figure~\ref{fig:utm_cont}$(b)$, to arrive at
    \begin{align}
        -\fot \int_{-\infty}^{\infty} \e^{-il (t-s)} \int_0^T \q(0,s) \ \d l \ \d s  =& \  \int_{\dD^-} \frac{ik \e^{-k^2 t} \big( \hf(k,T) + \v_o \big)} {(\alpha - k^2 + ik \gamma)} \ \d k\\
        & - \int_{\dD^-} ik \frac{ \e^{k^2 (T-t)} \big( \q(0,T) + \gamma \hq(k,T) \big)}{(\alpha - k^2 + ik \gamma)}  \ \d k, \nonumber
    \end{align}
    \begin{align}
        -\pi \q(0,t)  =& \  \int_{\dD^-} \int_0^t \frac{ik \e^{-k^2 (t-s)} \f(s)} {(\alpha - k^2 + ik \gamma)} \ \d k \ \d s + \int_{\dD^-} \frac{\v_o ik \e^{-k^2 t}} {(\alpha - k^2 + ik \gamma)} \ \d k  \nonumber \\
        &\ + \int_{\dD^-} \int_t^T \frac{ik \e^{k^2 (s-t)} \f(s) } {(\alpha - k^2 + ik \gamma)} \ \d k \ \d s \nonumber \\
        & - \int_{\dD^-} ik \frac{ \e^{k^2 (T-t)} \big( \q(0,T) + \gamma \hq(k,T) \big)}{(\alpha - k^2 + ik \gamma)}  \ \d k, \label{eq:gotoq}
    \end{align}
    where we have used the standard Fourier inversion formula on the left-hand side. There are two poles due to the quadratic in the denominator of above equation which are located at
    \begin{equation}
        k_\pm = \fot (i\gamma \pm \sqrt{4\alpha - \gamma^2}). \label{eq:pole}
    \end{equation}
    Since both $\gamma$ and $\alpha$ are positive parameters, the poles $k_\pm$ always lie in $\C^+$ and reach the real line for $\gamma = 0$, i.e., when Basset history is absent (see figure~\ref{fig:poles}). An appeal to Jordan's lemma~\citep[page 222]{ablowitz2003complex} shows that the third and fourth term (those involving function evaluations at $T$) in equation (\ref{eq:gotoq}) do not contribute, since $k^2$ is negative in $\D^-$ and $(t-s) > 0$. It is not surprising that these two terms do not contribute as this only ensures causality: the dynamics of the particle for all times $t < T$ cannot depend on a quantity from future time $T$. Using the fact that $\e^{-k^2t}$ is decaying in the region outside $\D^-$, we can write the final solution expression as:
    \begin{align}
        -\pi \q(0,t)  =& \ \int_{\dD^-} \int_0^t \frac{ik \e^{-k^2 (t-s)} \f(s)} {(\alpha - k^2 + ik \gamma)} \ \d s \ \d k - \int_{-\infty}^{\infty} \frac{\v_o ik \e^{-k^2 t}} {(\alpha - k^2 + ik \gamma)} \ \d k.
        \label{eq:solfin}
    \end{align}	
    Note that the particle velocity is now given by the above expression directly (and explicitly) in terms of the initial condition $\v_o (=\v(0)-\u(0))$  and the forcing $\f(t)$ in the relative frame of reference. The quantity $\q(0,t)$ represents the Dirichlet boundary condition for the modified Robin boundary value problem for the diffusion equation. Once the relative velocity $\q(0,t)$ is available, one may readily compute particle trajectories $\y(t)$ from (\ref{eq:BCposition}).

    Without the Unified Transform Method, one may not have suspected such a boundary value problem is indeed solvable. Of course, one could in principle have employed the Laplace transform to obtain the solution though a couple of issues arise then. If one were to use Laplace transform, the inverse would involve branch-cuts in the general case, which are difficult to evaluate, and as mentioned above, explicit solutions are most often not possible. However the Unified Transform Method provides the appropriate re-parameterisation to avoid these complications, and to further provide an explicit expression in $t$ for the solution, given by equation~(\ref{eq:solfin}).  As we shall see in the following sections, especially when considering the nonlinear problem $\f(\q(0,t),\y,t)$, the true benefit of adopting the new formulation is the local nature of the extended system which is precisely due to the connection between the MR equations and the diffusion equation.

    \section{Explicit solution for particular flows}
    In several fluid scenarios, there is a considerable separation between diffusive and convective time-scales. Two such examples are highlighted below. In such scenarios, the Basset history force has a non-negligible effect on the resultant particle dynamics. Consequently, we present several example flows which permit explicit treatment of the Basset history term and its impact on the particle motion.
    \begin{itemize}
    \item Clouds: In clouds the diffusion time-scale, $\tau$ for a droplet of size $a \sim 10 \mu$m is $\O$(ms) while the energy dissipation rate of $\varepsilon \sim 10 \rm{m}^2/s^3$ leads to a Kolmogorov eddy of size $\eta_k \sim 0.1$mm. Calculating the convective velocity scale $u_k$ using the relation $u_k \eta_k \sim \nu$ where $\nu$ is the kinematic viscosity, we get $u_k \sim 1$cm/s. From this we get the convective inertial time scale $\tau_{in} \sim \nu/u_k^2 = \O$(0.1s). We have a factor of $\O(100)$ showing that there is a long time region even at the length scale of a Kolmogrov eddy at which Basset history is expected to play a role.
    \item Planktons: Planktons in the ocean come in various sizes and different density ratios. The effect of turbulence on clustering~\citep{guseva2016history} is poorly understood and we believe Basset history, given its non-trivial dynamics, might play a role in its settling. A plankton of size $a \sim 100 \mu$m has a terminal velocity of $U_t \sim 0.1$mm/s when $\Delta \rho /\rho \sim 0.1$ (where $\Delta \rho$ is the difference between particle density and fluid density and $\rho$ being density of water). The diffusive time scale $a^2/\nu \sim 1$s while the $\tau_{in} = \nu/U_t^2 \sim \O$(hrs). This clearly shows that such a huge time scale separation makes studying Basset history effects worthwhile.	Also we suspect that the clustering of plankton could happen in short time, dynamics dominated by diffusive effects.
	\end{itemize}
	Given this motivation, we start looking at simple scenarios where fluid velocity is only time dependent and space independent as we build onto complex flows.
	
    \subsection{Example 1: A relaxing particle}
    \begin{figure}
        \centering
        \includegraphics[scale=0.9]{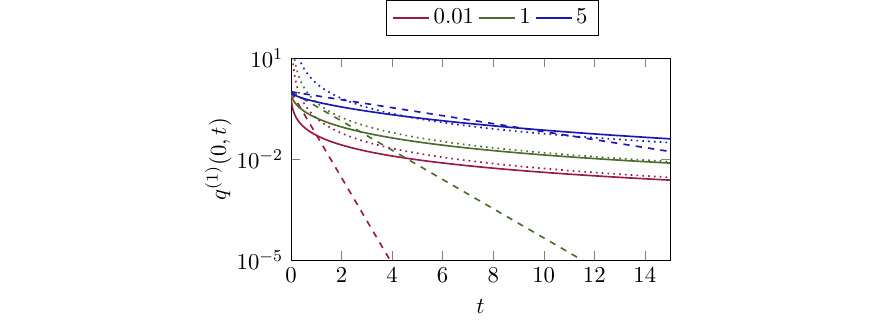}
        \caption{Particle velocity with (solid lines), and without (dashed lines) Basset history, along with the asymptotic solution (dotted lines) from equation (\ref{eq:asymp}).  The curves from bottom to top correspond to three different density ratios respectively: light particles, with $\beta(=\varrho_p/\varrho_f)=0.01$, neutrally buoyant particles with $\beta=1$, and heavy particles with $\beta=5$ at a fixed Stokes number of $S=1$. The decay rate of particle velocity goes as $t^{-3/2}$ for long times when Basset history is accounted for, in contrast to the exponential decay when it is neglected.}  
        \label{fig:zeroBG}
    \end{figure}    
    In the absence of a background flow (namely when $\f=0$), expression (\ref{eq:solfin}) for the relative velocity $\q(0,t)$, may be analysed  using Watson's lemma. Since the real-line in $k$ happens to be the path of steepest descent,  Laplace's method provides the asymptotic behaviour for functions of the form
    \begin{align}
        \mathscr{K}(\zeta):=& \int_{a}^{b} \e^{g(\zeta) t} \psi (\zeta)\ \d \zeta \ \text{as} \ t \rightarrow \infty, \ t > 0,
    \end{align}
    in terms of the local maximum value of function $g(\zeta)$ when $\zeta \in [a,b]$. This leads to the well-known asymptotic result for integrals of the above form
    \begin{align}
        \mathscr{K}(\zeta):=& \int_{-\infty}^{\infty} \e^{-\zeta^2 t} \psi(\zeta)\ \d \zeta \ \approx \  \sqrt{\frac{\pi}{t}} \sum_{n=0}^{\infty} \frac{\psi_{k}(0)}{2^{2n}n!}\ t^{-n}, \ \text{as} \ t\to\infty.
    \end{align}
    On setting $\f =0$  we see that the dominant contribution of the integral in (\ref{eq:solfin}) comes from $k=0$. To leading order we have
    \begin{align}
        \q(0,t) \approx& \frac{\v_o}{\sqrt{\pi} t^{3/2}} \frac{\gamma}{2 \alpha^2} + \O (t^{-5/2}), \ \quad \ t \gg 0.
        \label{eq:asymp}
    \end{align}
    \cite{farazmand2015,langlois2015} approached this problem by using Laplace transform. Their complete solution was provided in terms of convolutions and not directly as we obtain in equation (\ref{eq:solfin}). For large times however, \cite{langlois2015} present the same equation as (\ref{eq:asymp}). Their solution for a relaxing particle is the only earlier closed form solution that we know of, which accounts for the initial conditions correctly. We contrast the complete solution (\ref{eq:solfin}) with the one obtained upon neglecting the Basset history integral in the original MR equation. This is equivalent to neglecting the Neumann term $\q_x(0,t)$ in equation (\ref{eq:bcMR}) and repeating the calculation of $\q(0,t)$. Alternatively we just set $\gamma=0$ in (\ref{eq:solfin}) to obtain 
    \begin{align}
        \pi \q(0,t)  =& \ \int_{-\infty}^{\infty} \frac{\v_o ik \e^{-k^2 t}} {(k^2 - \alpha)} \ \d k.
    \end{align}
    The  non-zero contribution to the integrand comes only from the poles as the integrand is odd leading to
    \begin{align}
        \q(0,t) =& \ \v_o \e^{-\alpha t}, \label{eq:woBH}
    \end{align}
    and is precisely the expression one gets by solving the Maxey-Riley equation directly without the history term (and $\f=0$).	

    In figure~\ref{fig:zeroBG} we plot the velocity obtained both by including the history integral and excluding it for three cases: neutrally buoyant ($\beta=1$), light ($\beta=0.01$) and heavy particle ($\beta=5$). Two important features are evident. First, the solution with Basset history settles not exponentially but as a power-law (in this case a very slow $t^{-3/2}$). Secondly, there exists a cross-over time prior to which, a particle evolving under the influence of the Basset history integral and Stokes drag, relaxes faster than the particle evolving only under Stokes drag. However, after the cross-over time the particle that evolves only under Stokes drag relaxes faster. This contradicts the popular notion that Basset history acts as an effective drag. Thus far, for this example, we have considered the effect of varying density ratios on the particle relaxation time. Note, varying the Stokes number does not lead to substantially different velocity profiles since the Stokes number may be scaled out of the solution by replacing $k \rightarrow \tilde{k}/\sqrt{S}$ and $t\to\tilde{t}S$. Particle velocities for Stokes number $S\neq 1$ may be obtained from those depicted in figure~\ref{fig:zeroBG} by suitably scaling time.

    \subsection{Example 2: Sedimenting particles}
    The dynamics of particles settling under gravity is relevant in industrial applications, for aerosol particles and droplets in the Earth's atmosphere, and for carbon sequestration by marine snow in the oceans. In this section a two-dimensional flow is considered, with a superscript $(\cdot)^{(2)}$ indicating the vertical component of the relevant vector, while $(\cdot)^{(1)}$ indicates the horizontal component.  When the particle is acted upon by a body force such as gravity, the vertical component of forcing function $\f$, $f^{(2)}(t)$, is a constant of magnitude $\sigma$. This allows us to evaluate the time-transform explicitly as
    \[
        \tilde{f}^{(2)}(\omega,T) = \int_0^T \sigma \e^{\omega s} \ \d s = \frac{\sigma(\e^{\omega T}-1)}{\omega}.
    \]
    The settling velocity of the particle is
    \begin{align}
        \pi q^{(2)}(0,t)  =& \ \int_{-\infty}^{\infty} \frac{ik \e^{-k^2 t} v^{(2)}_o} {(\alpha - k^2 + ik \gamma)} \ \d k + \int_{-\infty}^{\infty} \frac{\sigma \gamma(1-\e^{-k^2 t})}{((\alpha - k^2)^2 + k^2 \gamma^2)} \ \d k.
        \label{eq:gravsol}
    \end{align}
    \begin{figure}
        \centering
        \includegraphics[scale=0.9]{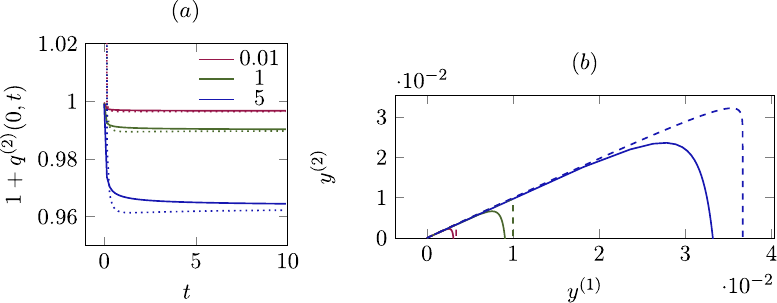}	
        \caption{$(a)$ Particle velocity relative to fluid, $q^{(2)}(0,t)$ along $y^{(2)}$-direction and $(b)$ particle trajectory as a function of time for three density ratios, $\beta = 0.01, 1, 5$ corresponding to light, neutrally buoyant and heavy particle when acted upon by gravity for $S=0.01, \sigma=-1$ in equation (\ref{eq:gravsol}). The asymptotic expression for velocity in (\ref{eq:gravasym}) is plotted as dotted line in $(a)$ and the dashed line in $(b)$ is the trajectory of particle without accounting for Basset history. Time taken by particles to reach $y^{(2)}=0$ when accounted for Basset history for these density ratios is $t=1.1, 1.15, 1.2$ and without Basset history it takes $t = 1, 1.01, 1.04$.}
        \label{fig:grav}
    \end{figure}	
    A solution for this problem is provided in \cite{clift2005}. However, their solution necessitates a zero initial-velocity condition for the particle. Moreover their solution is in terms of sums of error functions, and so the asymptotic form for large time is difficult to glean directly.  We evaluate the full solution (\ref{eq:gravsol}), without the zero initial-velocity requirement, for the same particle density ratios as in the previous example (but fixed Stokes number $S=0.01$), corresponding to neutrally buoyant, light and heavy particles: $\beta=1, 0.01$ and $5$, respectively, see figure~\ref{fig:grav}$(a,b)$. We also obtain the leading order asymptotic behaviour for the particle velocity 
    \begin{align}
        q^{(2)}(0,t) \approx& \ c(\alpha,\gamma) + \frac{\sigma \gamma}{\alpha^2 (\pi t)^{1/2}} + \O (t^{-3/2}), \quad \ t \gg 0, \\
%        c(\alpha,\gamma) =& \ \frac{\sigma}{\alpha} + \frac{2 i \gamma \sigma}{\sqrt{-\frac{1}{2} \sqrt{\gamma ^4-4 \alpha  \gamma
%                ^2}+\alpha -\gamma ^2/2} \sqrt{\sqrt{\gamma ^4-4 \alpha  \gamma
%               ^2}+2 \alpha -\gamma ^2}} \nonumber \\
%      & \times \frac{1}{\left(\sqrt{-\sqrt{\gamma ^4-4 \alpha  \gamma
%                ^2}+2 \alpha -\gamma ^2}+\sqrt{\sqrt{\gamma ^4-4 \alpha  \gamma ^2}+2
 %           \alpha -\gamma ^2}\right)}, \label{eq:gravasym}
 c(\alpha,\gamma) =& \ \int_{-\infty}^{\infty} \frac{\sigma \gamma} {((\alpha - k^2)^2 + k^2 \gamma^2)} \ \d k = \frac{\pi \sigma}{\alpha}, \label{eq:gravasym}
    \end{align}
    where $c(\alpha,\gamma)$ is the velocity of the sedimenting particle at long times. 

    Note that the terminal velocity $c(\alpha,\gamma)$ is in fact the same quantity as that obtained for a particle in the absence of the Basset history force and more importantly, this terminal velocity is independent of $\gamma$. The expression $\sigma\pi/\alpha$ is the constant velocity obtained from a balance of gravitational forces and steady drag in our notation. The evaluation of the integral expression for $c(\alpha,\gamma)$ is via an appeal to the residue theorem, albeit for three different cases: $\gamma^2/\alpha<4$, $\gamma^2/\alpha=4$ and $\gamma^2/\alpha>4$. In all three cases, the integral has the same value.

    We observe that a sedimenting particle attains its terminal velocity at a rate of $t^{-1/2}$, which is far slower than the exponential trend as predicted without Basset history. It is worth noting that a sedimenting particle's relaxation is even slower than the $t^{-3/2}$ trend observed for the force-free relaxation (\ref{eq:asymp}). Another interesting feature of this dynamics, which would be missed on neglecting the Basset history integral, is that the long time trajectory of the particle is given by $y^{(2)}\sim -1/(y^{(1)}-\xi)$ ($\xi$ being the final horizontal coordinate). This behaviour is anomalous in the sense that it is neither projectile nor ballistic. 

    The previous remarks suggest that Basset history could play a significant role in complex flows. For example, sedimentation rates of individual particles will determine how frequently particles collide with one another and coalesce as they sink. This question is of importance to estimating carbon sequestration by plankton in the ocean, raindrop growth, and for pulverised coal in thermal power plants. To this end, we make here a simple estimate for how a particle grows in size as it sediments through a sea of smaller particles.

    \begin{figure}
        \centering
        \includegraphics[scale=0.95]{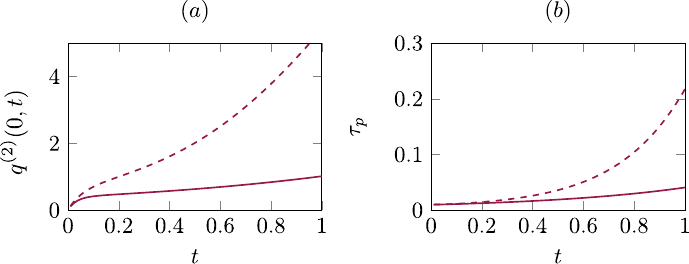}
        \caption{$(a)$ Velocity as a function of time, of a particle growing in size as it sediments through a uniform distribution of smaller particles and coalesces with those it collides with. $(b)$ Size of this particle as a function of time in terms of its relaxation time scale $\tau_p$ with initial $\tau_p=0.01 s, v^{(2)}_o=0.01, \nu = 10^{-5} m^2/s$ and  $\beta=10$ in an environment with solid fraction $\varphi=0.002$. Solid lines correspond to dynamics accounting for history effects while dashed to particles relaxation with Stokes drag alone.}\label{fig:dropgrow}
    \end{figure}

    It is simpler in this case to work with the dimensional equation, obtained by replacing $\sigma \rightarrow (\beta-1)g/R, S \rightarrow \tau_p$, where $g$ is gravity and $\tau_p=a^2/\nu$ is the particle time scale, so (\ref{eq:gravsol}) reads as
    \begin{align}
        \pi q^{(2)}(0,t)  =& \ \int_{-\infty}^{\infty} \frac{ik \e^{-k^2 t} v^{(2)}_o} {(\frac{1}{R \tau_p} - k^2 + \frac{ik}{R \sqrt{\tau_p}})} \ \d k - \int_{-\infty}^{\infty} \frac{\frac{(\beta-1)g}{R^2 \sqrt{\tau_p}} (1-\e^{k^2 t})}{((\frac{1}{R \tau_p} - k^2)^2 + \frac{k^2}{R^2 \tau_p})} \ \d k.
        \label{eq:gravsed}
    \end{align}
    Now $k$ has the dimensions of $t^{-1/2}$ and $v^{(2)}_o, q^{(2)}(0,t)$ are dimensional velocities. We follow one large particle of growing radius $a$ as it sediments through a quiescent fluid with a uniform distribution of identical small particles of radius $(a_s<a)$ occupying a solid fraction $\varphi$. The larger particle falls faster than the smaller ones, and in a short time $\Delta t$, it collides with all small particles whose centres lie within a cylinder of radius $(a+a_s)$ and height $q^{(2)}(0,t) \Delta t$. Assuming that all small particles it collides with will coalesce with the bigger particle and that $a \gg a_s$, its volume increases by $\varphi \pi a^2 q^{(2)}(0,t) \Dt$. Thus we get an evolution for the particle time-scale as
    \begin{equation}
        \tau_p^{(n)} = \bigg( \tau_p^{3/2} + \frac{3}{4} \varphi \tau_p \frac{q^{(2)}(0,t) \Dt}{\sqrt{\nu}}  \bigg)^{2/3},
    \end{equation}
    which modifies $\alpha, \gamma$ in equation~(\ref{eq:solfin}). After each time interval $\Dt$ we reinitialise the system with a new initial condition $v^{(2)}_o$ being the velocity from the previous time instant and a new relaxation time $\tau_p^{(n)}$. For a droplet of $\nu \sim 10^{-5} m^2/s,\ a \sim 0.3 mm, \beta=10$ we get $\tau_p \sim 10^{-2} s$. In figure~\ref{fig:dropgrow}$(a)$ we plot the dynamics of such a particle and in figure~\ref{fig:dropgrow}$(b)$ the particle size is represented in terms of its time scale $\tau_p$ as a function of time for two scenarios:  $(i)$ accounting for history effects (solid lines) and $(ii)$ only Stokes drag (dashed lines). We see that by neglecting the history force we will grossly overestimate particle size at a given time. 

    We have considered only vertical trajectories in this example. However in gravitational setting in turbulent particulate flows the anomalous trajectory could introduce non-trivial changes in the dynamics, which will be a topic for future study.

    \subsection{Example 3: Particle in an oscillatory background}
\begin{figure}
        \centering
        \includegraphics[scale=0.9]{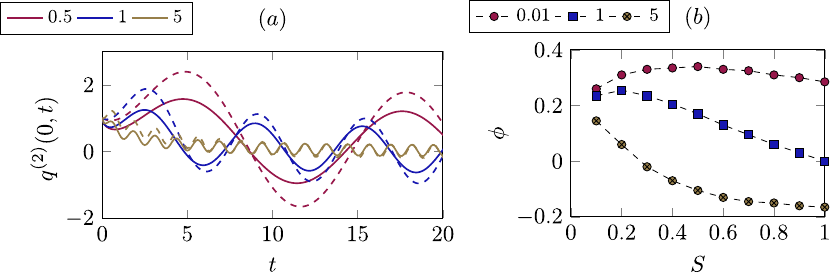}
        \caption{($a$) Particle velocity relative to oscillating background flow when $\beta=5$, $S=1$ for three different oscillating frequencies, $\lambda = 0.5, 1, 5$. The dashed lines correspond to the solution with Stokes drag alone, and the solid line accounts for Basset history as well, from equation (\ref{eq:osc}). ($b$) Phase difference, $\phi$ between solutions including and excluding history effects as a function of Stokes number, $S$, for a fixed $\lambda=1$ and three density ratios, $\beta=0.01, 1, 5$.}
        \label{fig:osc}
    \end{figure}

    We now consider the  behaviour of a single particle with a background flow that oscillates with a single frequency.  The forcing function is $f^{(2)}(s) =~ \sin (\lambda s)$ and the first term in equation (\ref{eq:solfin}) can be rewritten as
    \begin{align}
        \I =& \ \int_{-\infty}^{\infty} \int_0^t \frac{ik \e^{-k^2 m} f^{(2)}(t-m)} {(\alpha - k^2 + ik \gamma)} \ \d m \ \d k, \ \text{with} \ m = (t-s), \\
        =& \ \int_{-\infty}^{\infty} \frac{i \lambda k} {(\alpha - k^2 + ik \gamma)} \bigg[ \frac{\lambda \e^{-k^2 t} - \lambda \cos (\lambda t) + k^2 \sin (\lambda t) }{(\lambda^2+k^4)} \bigg] \ \d k. \label{eq:osc}
    \end{align}
    The solution for the scenario with Stokes drag (but no Basset history integral) is
    \begin{align}
        q^{(2)}(0,t) =& \ q^{(2)}(0,0) \e^{- \alpha t} +  \frac{\lambda \e^{- \alpha t}}{(\alpha ^2+\lambda ^2)} + \frac{\alpha  \sin (\lambda t)- \lambda  \cos (\lambda  t)}{(\alpha ^2+\lambda ^2)}.\label{eq:oscStokes}
    \end{align}
    We evaluate this velocity expression as a function of time in figure~\ref{fig:osc}$(a)$ and find that with Basset history, particles oscillate with smaller amplitudes at early time, and that they attain their final periodic state faster than when Basset history is neglected. This behaviour, governed by the short-time dynamics, is qualitatively different from that of the cases we studied in earlier sections, where Basset history significantly slowed down the attainment of terminal velocity. In particular, the Basset history integral is not simply an additional drag. We see that expressions (\ref{eq:osc}) and (\ref{eq:oscStokes}) asymptote to similar expressions, albeit with a phase difference.
    We can quantify the phase difference $\phi$ that persists at long times between a particle with history effects and with only Stokes drag at long times. This is shown in figure~\ref{fig:osc}$(b)$ for Stokes number in the range $S=0.1-1$ and for three different density ratios, $\beta = 0.01, 1, 5$. We should however remind ourselves that though we have chosen $\lambda \sim \O(1)$ where particle relaxation time scale and forcing are similar, the solution expression is general and applies to all scenarios with frequencies capped by the diffusive time scale.

    \section{Spatially dependent flows}
    Thus far we considered the MR equation with spatially uniform fluid flows. When the fluid flow varies in space, the MR equation is generically nonlinear and one does not obtain a closed-form expression for the particle velocity $\q(0,t)$. Most fluid flows are indeed spatially dependent. The role of the Basset history integral in determining the dynamics of inertial particles in such fluid flows has been investigated by others (\cite{olivieri2014effect,guseva2017snapshot,daitche2011memory,daitche2014memory,daitche2015role}). These studies focused on the contribution of the Basset history to particle collision rates, preferential concentration and residence time of heavy/light particles. All of the aforementioned studies employed the numerical scheme developed in \cite{daitche2013} which approximates the singular kernel of the Basset history integral.  In this section, we show how our reformulation of the MR equation may be readily adapted to the case of spatially dependent fluid flows. We emphasize that no approximations are made in handling the singular kernel. Furthermore, the memory costs are constant in time.

    We begin the discussion in subsection \ref{couette_section} by considering the dynamics of a particle in planar Couette flow. Since the fluid velocity is linear in the spatial variable, this spatially dependent problem is in fact exactly solvable. Moreover the exact solution found here may be used to simulate the dynamics of particles in more complicated, but slowly varying, flow fields by using the the leading order term in a gradient expansion of the fluid velocity.

    Our next example, in \ref{vortex_section}, is that of a particle in the neighbourhood of a point vortex. Here we probe the effect of the history-integral on caustics formation and also the large-time behaviour of the particle. This requires some modification of our original method to handle nonlinear forcing functions, which we describe in detail below. The extension of the numerical method, presented in this section, to cases where the fluid velocity is not specified explicitly is straightforward, though not the topic of the present manuscript. It is thus our contention that the numerical method we develop here is directly applicable for including the history integral in any general flow.

    \subsection{Example 4: Migration in plane Couette flow}
    \label{couette_section}
    The Couette flow velocity profile is given by $u^{(1)} = \lambda y^{(2)}$. As before, the superscripts $(\cdot)^{(1)}$ and $(\cdot)^{(2)}$ represent the horizontal and vertical components respectively. The particle velocity evolution in such a background flow is given by
    \begin{align}
        \partial_t
        \begin{bmatrix}
            q^{(1)}(0,t) \\
            q^{(2)}(0,t)
        \end{bmatrix} = & \ 
        \underbrace{
            \begin{bmatrix}
                \alpha & \lambda \\
                0 & \alpha
            \end{bmatrix}}_{\M}
        \begin{bmatrix}
            q^{(1)}(0,t) \\
            q^{(2)}(0,t)
        \end{bmatrix}
        +
        \gamma
        \underbrace{
            \begin{bmatrix}
                1 & 0 \\
                0 & 1
            \end{bmatrix}}_{\mathbb{I}}
        \begin{bmatrix}
            q^{(1)}_x(0,t) \\
            q^{(2)}_x(0,t)
        \end{bmatrix}.	
    \end{align}
    \begin{figure}
        \centering
        \includegraphics[scale=0.9]{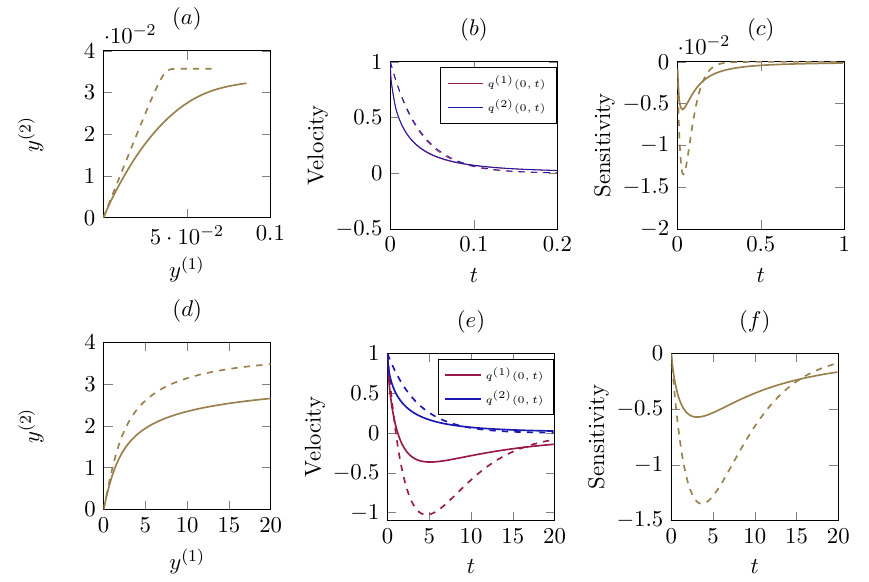}	
        \caption{Particle dynamics in Couette flow: solid lines are the solutions in equation (\ref{eq:PCflow}), while dashed lines indicate dynamics upon neglecting Basset history in equation~(\ref{eq:PCwo}). Top row corresponds to solution for $S=0.01$ and bottom for $S=1$ for a fixed density ratio of $\beta=5$. $(a, d)$ Particle trajectory $y^{(2)}$ as a function of time, ($b, e$) Horizontal component $q^{(1)}(0,t)$ and vertical component $q^{(2)}(0,t)$ respectively of particle velocity relative to fluid velocity with initial velocity of particle relative to fluid $\q(0,0)=1$. $(c, f)$ Off-diagonal component of sensitivity matrix in equation (\ref{eq:sens}) which is due to shear in the flow.}
        \label{fig:PCflow}
    \end{figure}	
    Since the boundary condition is linear with constant coefficients, we may follow a similar procedure as earlier. We now have two diffusion equations coupled at an interface where the interfacial boundary condition in each is given by the MR equation~(\ref{eq:BCmod}). Note however, that the diffusion equations themselves (i.e. in the bulk) are decoupled from each other. As a result, the global relation for this system of diffusion equations is just the vector version of (\ref{eq:gr})
    \begin{equation}
        \e^{\omega(k) T} \hat{\q} (k,T) + \vg_1(k,T) + ik \vg_o(k,T) = 0, \ \quad \ k \in \C^-.
    \end{equation}
    The boundary condition becomes
    \begin{align}
        \bigg( \q(0,T) \e^{\omega T} - \q(0,0) \bigg) -k^2 \vg_o(k,T) = & \ -\M(\alpha,\gamma,\lambda) \vg_o(k,T) + \gamma \vg_1(k,T).
    \end{align}
    Eliminating $\vg_1(k,T)$ just as earlier from the above equation, we get for $\vg_o(k,T)$
    \begin{align}
        \big( (k^2 -ik\gamma) \mathbb{I} -\M(\alpha,\gamma,\lambda) \big) \vg_o(k,T) =& \   \bigg\{ \q(0,T) \e^{\omega T} - \q(0,0) + \gamma \hat{\q}(k,T) \e^{\omega T} \bigg\} \\
        \mbox{ where }\big( (k^2 -ik\gamma) \mathbb{I} -\M(\alpha,\gamma,\lambda) \big)^{-1} = & \
        \begin{bmatrix}
            (k^2 - \alpha - ik \gamma)^{-1} & \lambda(k^2 - \alpha - ik \gamma)^{-2} \\
            0 & (k^2 - \alpha - ik \gamma)^{-1}
        \end{bmatrix}.
    \end{align}

    Multiplying the above equation by $ik\e^{-k^2t}$ and integrating over $\dD^-$ we obtain the solution 
    \begin{align}
        \pi \q (0,t)
        =& \ -\int_{-\infty}^{\infty} ik \e^{-k^2t}
        \begin{bmatrix}
            (k^2 - \alpha - ik \gamma)^{-1} & \lambda(k^2 - \alpha - ik \gamma)^{-2} \\
            0 & (k^2 - \alpha - ik \gamma)^{-1} \label{eq:PCflow}
        \end{bmatrix} \q(0,0) \ \d k.
    \end{align}
    On the other hand, the solution to the MR equation without the Basset history integral simplifies to
    \begin{align}
        \q(0,t) =& \  \e^{-\alpha t}
        \begin{bmatrix}
            1 & -\lambda t \\
            0 & 1
        \end{bmatrix}
        \q(0,0).\label{eq:PCwo}
    \end{align}
    The results are provided in figure~\ref{fig:PCflow}. Sub-figures $(a)$ and $(d)$, respectively for Stokes number  $S=0.01$ and $1$, show particle trajectories with and without Basset history. The $y^{(2)}$ coordinate has saturated to the values shown. Clearly the migratory behaviour is different, and neglecting Basset history predicts the final locations of the particles incorrectly. Furthermore, the relaxation time is longer with Basset history, i.e. the velocity decay is slower, as seen in  figure~\ref{fig:PCflow}$(b,e)$. Note that the relative velocity is plotted here, so each particle at long time attains the local velocity of the Couette flow ($=\lambda y^{(2)}$). The solution expressions (\ref{eq:PCflow}-\ref{eq:PCwo}) readily lead to the following expressions for the sensitivity of the solution to a particle's initial velocity
    \begin{align}
        \bigg(\frac{\delta \q(0,t)}{\delta \q(0,0)} \bigg)_{\text{His}} =& \ 
        -\frac{1}{\pi}\int_{-\infty}^{\infty} ik \e^{-k^2t}
        \begin{bmatrix}
            (k^2 - \alpha - ik \gamma)^{-1} & \lambda(k^2 - \alpha - ik \gamma)^{-2} \\
            0 & (k^2 - \alpha - ik \gamma)^{-1}
        \end{bmatrix} \ \d k, \\
        \bigg(\frac{\delta \q(0,t)}{\delta \q(0,0)} \bigg)_{\text{Sto}} =& \  \e^{-\alpha t}
        \begin{bmatrix}
            1 & -\lambda t \\
            0 & 1
        \end{bmatrix}.\label{eq:sens}
    \end{align}
    The eigenvalues of this sensitivity matrix provide the Lyapunov exponents and both eigenvalues are equal to $1$ for the scenarios with and without Basset history. However the eigenvectors corresponding to these matrices are different and in figure~\ref{fig:PCflow} $(c,f)$ we plot the evolution of the off-diagonal component for two different $S=0.01, 1$.

     \subsection{Solution for non-linear spatial flows}
    \label{nonlin}	
    So far, we have been able to explicitly compute the particle velocity, from $\hg_0(\omega,T)$, see equation (\ref{eq:gotoq}). However when the spatial dependence of the fluid is flow is nonlinear, i.e., when $\f$ depends on $\y$, equation (\ref{eq:gotoq}) is not an explicit expression for the particle velocity. Consequently we resort to a numerical scheme to evolve the dynamics of the particle, which we present in this section. Our numerical method has the following advantages:
    \begin{itemize}
        \item The method completely eliminates the need to store the entire trajectory of the particle to evaluate the Basset history term.
        \item Our method achieves spectral accuracy by employing Chebyshev polynomials to evaluate the integrals involved.
        \item The method involves a fixed memory cost (of $\O(N)$, $N$ being number of Chebyshev modes) independent of time, making long time simulations no more expensive than short time. This is useful to compute statistics in turbulent simulations.
        \item The effect of history can be recorded (with fixed memory size), enabling simulations to be restarted. Such a restart is impossible in other techniques owing to exponentially large memory costs in evaluating the integral. While preparing the manuscript we became aware of the work of \cite{parmar2018} who approximate the history kernel by a summation of exponentials and achieve the desired accuracy. However we would like to highlight that our technique does not make any approximation to the kernel and we show later that all information due to history can be stored in a variable whose computation is of fixed memory throughout particle evolution. Moreover the form of the solution in equation (\ref{eq:solfin}) is specially useful for studying asymptotic behaviour (ref.~\cite{vishal2018}), unlike other techniques.
    \end{itemize}

    In the following subsection we derive the equations we ultimately solve for the particle trajectory and in the subsequent subsection we provide the details of the numerical implementation.

    \subsubsection{Evolution equation for Basset history-integral}
    \begin{figure}
        \centering
        \includegraphics[scale=0.85]{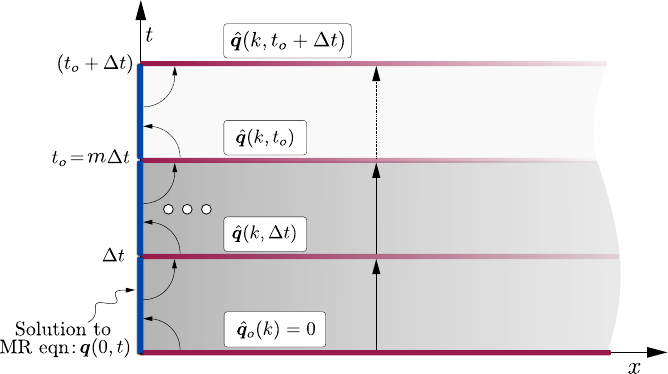}
        \caption{Schematic showing numerical scheme where we evaluate $\hq(k,t)$ at intervals $t_o, (t_o+\Dt)$ from the solution at $t \in [t_o-\Dt,t_o], [t_o,t_o+\Dt]$ and so on in order to eliminate the need to store $\q(0,t)$ from $t \in [0,t_o+\Dt]$.}
        \label{fig:schmNM}
    \end{figure}
    We begin by dividing the $x-t$ plane into domains of size $[0,\infty)\times[t_o-\Dt,t_o]$, where $t_o=m\Dt$ for some integer $m$, as shown in figure~\ref{fig:schmNM}. Our procedure involves two steps, repeated for each such domain:
    \begin{enumerate}[(a)]
        \item \ Using the initial condition $\q(x,t_o-\Dt)$ in each domain (bottom wall of that domain in the figure), we compute the boundary condition $\q(0,t), t \in (t_o-\Dt, t_o]$ (left wall of that domain). This is done by solving the Maxey-Riley equation, in the form given by equation (\ref{eq:fin1}) below. This equation is the extension of (\ref{eq:solfin}) to the case of arbitrary initial conditions. Moreover the forcing $\f$ is now a function of $\q(0,t)$ and/or $\y(t)$. So we start with a guess for $\q(0,t)$ and $\y(t)$, and iterate using a nonlinear solver. We thus have the Dirichlet condition on the left wall of figure~\ref{fig:schmNM}, over $[t_o-\Dt,t_o]$.
        \item \ Using the computed boundary condition $\q(0,t)$ from the previous step and the known initial condition $\q(x,t_o-\Dt)$, we directly compute $\q(x,t_o)$, the top wall of that domain, by using equation (\ref{eq:fin2}) which will be derived below. This step is equivalent to solving the diffusion equation with a Dirichlet boundary condition and the solution of this boundary-value problem leads to the initial condition for the next domain, $t \in (t_o, t_o+\Dt]$. Through an efficient use of the global relation, we do not require the explicit solution expression $\q(x,t_o)$, but only the Fourier transform $\hq(k,t_o)$ which simplifies the calculations considerably.
    \end{enumerate}
We recognise that our system size (namely, the number of dependent variables) is higher for a nonlinear case, and this is the price we need to pay to eliminate the rising memory costs coming from the nonlocal time integral term.

    In the first domain alone $[0,\infty)\times[0,\Dt]$, equation (\ref{eq:solfin}) may be used to obtain the particle velocity. We recognise that $\f$ now depends on $\y$ and $\q(0,t)$. Along with (\ref{eq:BCposition}), where we set $T=\Dt$, we now have two equations for two unknowns $\y$ and $\q(0,t)$ over the time interval $[0,\Dt]$. In a standard numerical method, one would approximate the first integral of (\ref{eq:solfin}) using a quadrature rule. We however, adopt an alternative route (partly to ensure accuracy) by representing the unknowns $\q(0,t)$ and $\y(t)$ in terms of Chebyshev polynomials over the interval $[0,\Dt]$. All integrals are readily evaluated to a high degree of accuracy. With the Chebyshev polynomial representation, equations (\ref{eq:solfin}) and (\ref{eq:BCposition}) are solved using a nonlinear solver, here Newton's method, for the coefficients of the Chebyshev polynomials. Once we have $\q(0,t)$ (in terms of Chebyshev polynomials), we solve the diffusion equation with this \emph{Dirichlet condition} to obtain the solution $\q(x,\Dt)$. The evaluation of such new initial conditions at every $(t_o = m \Dt)$ is possible precisely because of the relationship between Basset history integral and Neumann-condition to diffusion equation. 

In all later domains, the initial condition $\q(x,t_o)$ is no longer zero. Indeed $\q(x,t_o)$ is the solution of the diffusion equation at this time. 
The global relation of the diffusion equation in the time interval $t \in (t_o,t_o+\Dt]$ can be written as
    \begin{align}
        \e^{k^2 (t_o+\Dt)} \hq (k,(t_o+\Dt)) =& \  \e^{k^2 t_o} \hq (k,t_o) - ik \hg_o(k^2, t_o,t_o+\Dt) - \hg_1(k^2,t_o,t_o+\Dt),  \ k \in \C^-. \label{eq:GRuseful}
    \end{align}
    For the boundary terms represented by $\hg_{i}(k^2,t_1,t_2)$, we have introduced the additional arguments $t_1, t_2$, which are the limits of the integral defining the time transform. This notation is also used in the forcing, as $\hf(k^2,t_1,t_2)$. Following a procedure similar to that employed to derive equation (\ref{eq:gotoq}), we obtain the following expression for $\q(0,t)$ 
    \begin{align}
        -\frac{\pi}{2} \q(0,t) = & \ \underbrace{\int_{0}^\infty  k \e^{-k^2(t-t_o)} \Im \{ \H(k,t_o) \} \ \d k}_{\His(t)} \nonumber \\ & + \underbrace{\int_{0}^\infty \Im \bigg\{ \frac{k \e^{-k^2t}\hf(k^2,t_o,t)}{(\alpha - k^2 + ik\ga)} \bigg\} \ \d k}_{\F(t)} ,\quad t \in (t_o, t_o+\Delta t], \label{eq:fin1}
    \end{align}
    The term $\H(k,t_o)$ in the integrand of the first integral in the expression above, is the modified initial condition  given by
    $$
    \H(k,t_o) = \bigg\{ \frac{\q(0,t_o)+ \gamma \hq(k,t_o)}{(\alpha - k^2 + ik \ga)} \bigg\}.
    $$
    This term encodes all the information about the history of the particle from $t \in (0,t_o]$. Equation (\ref{eq:fin1}) is thus the extension of equation (\ref{eq:solfin}) for a general initial condition to the diffusion equation. The renormalised initial condition $\H(k,t)$ itself evolves in time and this evolution is once again obtained from the global relation, see appendix \ref{ssec:histevol} for details. We state the relevant equation here.
    \begin{align}
        \H(k,t_o) =& \ \e^{-k^2\Delta t} \H(k,t_o - \Delta t)  -  \underbrace{\e^{-k^2 t_o} \int_{t_o - \Delta t}^{t_o} \e^{k^2 s}\q(0,s) \ \d s}_{\I_1} -  \underbrace{\e^{-k^2 t_o} \frac{\hf(k^2,t_o - \Delta t, t_o)}{(\alpha - k^2 + ik\gamma)}}_{\I_2}.
        \label{eq:fin2}
    \end{align}
    Note all terms on right-hand side of (\ref{eq:fin2}), including the forcing term, are known since they depend on the previous time interval $[t_o-\Dt,t_o]$. Equation (\ref{eq:fin2}) may also be written as a dynamical system for individual modes of history as
    \begin{align}
        \dot{\H}(k,t) + k^2 \H(k,t) =& \ \q(0,t) + \frac{\f(t,\y(t),\q(0,t))}{(\alpha - k^2 +ik\ga)},  \ t \in [t_o-\Delta t, t_o].
    \end{align}
    Equations (\ref{eq:fin1}) and (\ref{eq:fin2}) represent the solution to the full MR equations for a particle in a time-dependent inhomogeneous flow field. Note the individual modes $k$ of $\H(k,t)$ in equation (\ref{eq:fin2}) evolve independently. However  the expression for the velocity in equation (\ref{eq:fin1}) depends on all of them, and indeed depends on the integral of the history. 

We are now enabled to march repeatedly till any $t_o$ from any $(t_o-\Dt)$ using the set of initial conditions $\hq(k,t_o-\Dt)$, solving equation (\ref{eq:fin1}) for $\y(t)$ and $\q(0,t)$ in the interval  $(t_o-\Dt)$ to $t_o$, and using these Dirichlet boundary conditions in equation (\ref{eq:fin2}) to get  $\hq(k,t_o)$. Once again, we emphasize that the above procedure eliminates the need to store the entire particle velocity from the initial time instant. The effect of the Basset history integral is encoded in $\H(k,t)$.

    \subsection{Numerical technique}\label{sec:NumMet}
    We now detail a numerical scheme  to solve equations (\ref{eq:fin1}-\ref{eq:fin2}). The continuous variables $t$ and $k$ are restricted to a set of discrete values over their respective intervals, $t\in (t_o-\Dt, t_o]$ and $k \in [0,\infty)$, leading to discrete times $t_j$ and discrete wavenumbers $k_l$ where $j=1, \dots, L; l=0, \dots, N$. In the case of $t$ we employ Chebyshev nodes under an affine transform whereas for $k$ we employ a mapped Chebyshev node appropriate for rational Chebyshev approximations. We then expand $\q(0,t)$ and $\H(k,t_o)$ in a Chebyshev polynomial basis $T_n(t)$ (rational Chebyshev polynomials for $\H$). We adopt a collocation approach to solve the resulting nonlinear equation. 

    The discrete version of equation (\ref{eq:fin1}) can be written as a nonlinear equation for the coefficients as
    \begin{align}
        \mathscr{T}(t_j) \equiv \frac{\pi}{2} \q(0,t_j) + \His(t_j) + \F(t_j)=0 , \ t_j =t_o + \frac{\Dt}{2} \bigg(1 + \cos \big(\frac{j\pi}{L}\big) \bigg), \label{eq:march}
    \end{align}
    where $j=1, \dots, L$. Details on how to compute each term in the above equation are presented in the Appendix~\ref{app:Ntech}. The contribution from $\H(k,t_o)$ is captured in $\His(t_j)$, whereas the contribution from the nonlinear forcing due to the fluid velocity field is present in $\F(t_j)$. An equation for the position, $\y$ can be obtained from the discretised version (\ref{eq:BCposition}). A standard Newton root finder is employed to solve for $\q(0,t)$ and $\y(t)$ in the time interval $(n\Dt, (n+1)\Dt]$ for each $n=0,1,\ldots$ with an appropriate initial guess, typically  $\q(0,t_j) = -2/\pi \His(t_j)$.  Equation (\ref{eq:fin2}) on the other hand is a straightforward integral evaluation using known values of  $\q(0,t_j)$ and $\H(k_j,t_o-\Dt)$ from the previous time step. We show in the appendix that expanding $\q(0,t)$ and $\H(k,t_o-\Dt)$ in Chebyshev basis allows us to pre-compute several  terms which speeds up evaluation of expressions in (\ref{eq:fin2}). 

    \begin{figure}
        \centering
        \includegraphics[scale=1.0]{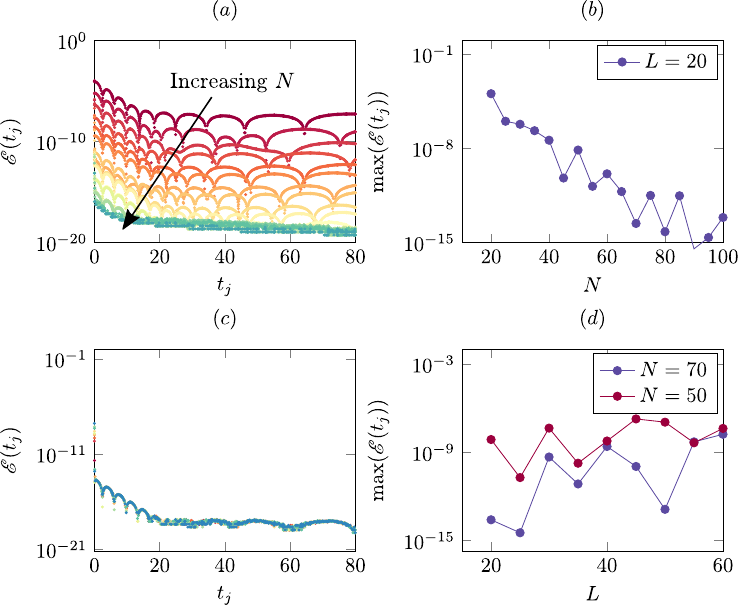}
        \caption{Error estimates for the solution to a relaxing particle using the numerical scheme developed in section~(\ref{sec:NumMet}). Error as a function of time for $(a)$ different values of $N$ at fixed $L=20$ and $(c)$ different values of $L$ at fixed $N=50, 70$. $(b), (d)$ Maximum of the error as we increase the number of Chebyshev modes in $k$ and $t$.}
        \label{fig:error}
    \end{figure}

    \subsection{Convergence properties and memory requirement of the numerical scheme}
    To study convergence of our numerical scheme, we compare the exact velocity of a relaxing particle (for which we have an analytic expression) to the numerical solution obtained from the scheme described above. We first check the numerical method's accuracy as a function of the number of Chebyshev nodes $N$ in wave number $\tk$, where $\tk : \ k \in [0,\infty]\rightarrow[-1,1]$. We also check the accuracy for number of Chebyshev nodes $L$ in $t \in [t_o,t_o+\Dt]$ within a time interval of length $\Dt$.	The error in the calculated solution is given by 
    \[
        \E(t_j) = | q^{(1)}(0,t_j) - q^{(1)}_e(0,t_j) |
    \]
    where $q^{(1)}_e(0,t_j)$ is the exact solution evaluated at discrete $t_j$ and $q^{(1)}(0,t_j)$ is the solution evaluated using the present numerical technique. In figure~\ref{fig:error}$(a)$ we plot the error as a function of time for all the number of $\tk$ chosen and we see its magnitude decrease with increase in $N$. In figure~\ref{fig:error}$(b)$ we plot the maximum value of $\E(t_j)$. We clearly see that error is small and decreases exponentially with increase in $N$ for fixed $L=20$. We can reach a desired accuracy by choosing the appropriate number of Chebyshev nodes $N$. Moving now to the convergence of solution with increase in nodes $L$ in $\Dt$ we see there is a general increase in error in figure~\ref{fig:error}$(d)$ for both fixed $N=50$ and $70$. Such a behaviour is expected and not an anomaly. Firstly observe that the maximum error is attained, for all different values of $L, N$, at the first grid-point i.e. $t_1$ in $t_j \in (0,\Dt]$. This is because of the $t^{-1/2}$  singularity at $t=0$. As we increase the number of Chebyshev nodes in the interval of length $\Dt$, the first Chebyshev grid-point moves closer to this singularity. We can easily account for this behaviour by increasing the number of modes in space $N$. We show this by increasing the spatial Chebyshev modes from $N=50$ to $70$ in figure~\ref{fig:error}$(d)$ which results in a drop in the maximum error.

    We emphasize the advantage a Chebyshev basis provides in terms of computational efficiency. In order to compute the Chebyshev coefficients via the Discrete Chebyshev Transform (\texttt{DCT}), we only require $\O(N \log N)$ operations where $N$ is the number of Chebyshev modes. This is done by leveraging Fast-Fourier Transform (\texttt{FFT}) package to evaluate the \texttt{DCT}s involved in the process of computing Chebyshev coefficients.

    The memory requirement for our technique comes from storing the variables $\q(0,t_j)$ and $\H(k_j,t_o)$. Considering $L$ modes in $t_j$, $N$ modes in $\tk_j$ demands only  storage of $\O(N+L)$ values for all times. Whereas the other techniques require a linearly increasing memory cost of $\O(t/\Dt)$. For $N=30, L=20$ one requires to store only $50$ values which remain fixed throughout the simulation. In this estimate we of course do not account for the pre-computed matrices which turn out to be three $N\times N$ matrices (see Appendix~\ref{app:Ntech}) resulting in another $2700$ values for $N=30$ but this initial cost remains constant and helps in shorter computation time.

    \subsection{Example 5: Single point vortex and caustics}\label{vortex_section}
    The stationary flow-field corresponding to a single vortex is
    \begin{align}
        \u =& \ \frac{\Gamma}{2 \pi} \ez \times \frac{\y(t)}{|\y|^2}.
    \end{align}
    Though this flow field does not have a time varying component, it still has contributions from both $(\u\cdot \nabla) \u$ and $(\q(0,t) \cdot \nabla) \u$. This makes the system of evolution equations for particle position, $\y(t)$ and velocity, $\q(0,t)$ coupled and non-linear. The solution is given by
    \begin{align}
        \y(t) =& \ \y(0) + \int_o^t \q(0,s) \ \d s + \int_o^t \frac{\Gamma}{2 \pi} \ez \times \frac{\y(s)}{|\y|^2} \ \d s,
    \end{align}
    for the velocity evolution
    \begin{align}
        \q_t(0,t) + \alpha \q(0,t) - \gamma \q_x(0,t) =& \ \bigg( \frac{1}{R} - 1 \bigg) (\u \cdot \nabla) \u - (\q(0,t) \cdot \nabla) \u.
    \end{align}
    \begin{figure}
        \centering
        \includegraphics[width=\textwidth]{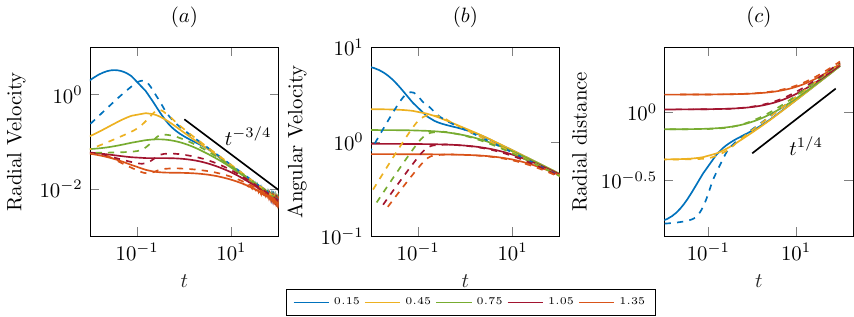}
        \caption{$(a)$ Radial velocity,  $(b)$ angular velocity and $(c)$ radial distance travelled by particles starting at 5 different locations separated by a non-dimensional distance of $0.15$ at $S=0.01, \beta=10$ and with initial velocity of $\v_o=(0.05,0.05)$, $\Gamma=2\pi$, the solid black lines correspond to power-law of $t^{-3/4}, t^{1/4}$. Solid lines represent simulation including the Basset history integral, Stokes drag and added mass effects while the dashed lines are for Stokes drag with added mass.}
        \label{fig:singleVor}
    \end{figure}
    Using the numerical technique described in the previous section, we solve for position and velocity for a Stokes number of $S=0.01$, density ratio $\beta = 20$ and plot the radial and angular velocity in figure~\ref{fig:singleVor}$(a,b)$. Different lines in the figure correspond to different initial position separated by a distance of $0.15$ though all particles start with a initial velocity of $(0.05,0.05)$. Once the particles get centrifuged far away from the vortex, the trajectory follows $y^{(r)}(t) \sim t^{1/4}$ as shown in figure~\ref{fig:singleVor}$(c)$, where $y^{(r)}$ is the radial distance from the point vortex. Thus the radial velocity, $q^{(r)}(0,t) \sim t^{-3/4}$ as in figure~\ref{fig:singleVor}$(a)$. This long-time behaviour is the same as that obtained by~\cite{ravichandran2015} without Basset history. The history integral affects the transient behaviour and the approach to the long-time limit.

\begin{figure}
        \centering
        \includegraphics[scale=0.9]{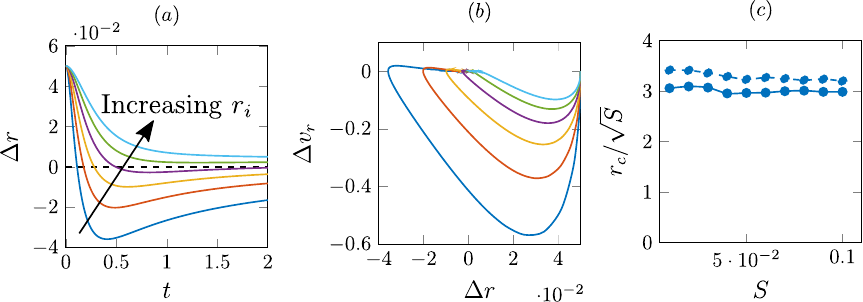}
        \caption{$(a)$ Evolution of the difference in radial distance between particles $\Delta r$ for different initial particle locations $r_i$ separated by a distance $0.05$ for Stokes number, $S=0.02$. The critical particle location at which $\min(\Delta r)$ crosses $t$-axis is the location of end of caustic formation. $(b)$ The phase-portrait defined by $\Delta v_r$ the difference in radial velocity vs $\Delta r$ also shows the transition from caustic formation to no caustic formation with increase in $r_i$. $(c)$ Critical radius $r_c$ for caustic formation as a function of Stokes number $S$ with history-integral, added mass (solid line) and with Stokes drag alone (dashed line). }
        \label{fig:Caust}
    \end{figure}

    We discussed in the introduction about caustics near a vortex. As shown by~\cite{ravichandran2015} analytically, and in two-dimensional simulations of turbulence, particles inside a critical radius, $r_c$ are evacuated rapidly, resulting in a region of high particle concentration in the neighbourhood of $r_c$. In order to identify this critical radius including the effects of Basset history, we track the radial separation $\Delta r$  between two particles that are initially placed close to each other. This is shown in figure~\ref{fig:Caust}$(a)$ for various initial radial distance $r_i$ from the vortex of the inner particle (the one closer to the vortex). When $\Delta r$ crosses zero, caustics have occurred. We find that caustics occurs for $r_i \le 0.44$, i.e., $r_c=0.44.$ for this Stokes number $S=0.02$. Also, the trajectories of particles that exhibit caustics take different paths compared to ones that do not, in the phase-plane defined by $\Delta v_r$ vs $\Delta r$ where $\Delta v_r$ is the difference in radial velocities for two closely placed particles (see~\cite{ravichandran2015}). This is shown in figure~\ref{fig:Caust}$(b)$ where a negative value of $\Delta v_r$ implies that the inner particle travels faster than its neighbour. When $\Delta r$ goes negative it has overtaken its neighbour. Thus trajectories that cross the $y-$axis of the phase-plane form caustics. 

    In figure~\ref{fig:Caust}$(c)$ we plot the critical radius, $r_c$ (found by the location where $\min(\Delta r(t))$=0) as a function of the Stokes number $S$. Simulations with history integral and added mass effects included are shown as solid lines and with just Stokes drag as dotted lines. The history integral and added mass together reduce the region over which caustics are formed, and may cause a reduction in the number densities of particles predicted without the history term. The implications for particle dynamics in turbulent fluids flows are beyond the realm of present article and will be explored in our upcoming study.

    \section{Discussion}
    \subsection{Interpretation of the Dirichlet-Neumann map}
    The map that we establish between the Basset history integral and the Neumann condition of the 1-D diffusion equation arises on mathematical grounds, but can also be interpreted physically as follows. The diffusion equation that the variable $\q(x,t)$ obeys can be seen as diffusion of momentum from the particle boundary and the Neumann condition $\q_x(0,t)$ on the other hand would mean the flux of momentum transferred from the particle boundary in the form of shear stress to the bulk of fluid. The trail left by the particle motion is a consequence of this momentum flux that emanates from the particle boundary which has been shown to arise from the map between Dirichlet-Neumann conditions to diffusion equation. This relation to the diffusion or heat equation further emphasizes the fact that our analysis accounts for the momentum diffusion valid before any convective inertial effects impact the `wake' generated by the particle. 
    
    \subsection{Different geometries}
    In the present manuscript, we have focussed on our reformulation of the Maxey-Riley equation for spherical particles. The technique used employed here is, however, general and extends immediately to other particle shapes. The only additional requirement for spheroids and disks travelling parallel to their axis of symmetry is the redefinition of the three non-dimensional quantities
    \[
        \hat{\beta} = \beta/\cC_A, \quad \hat{\alpha} = \cC_a \alpha, {\rm and} \quad \hat{\gamma} = \cC_H \gamma,
    \]
    which respectively appear in equation~(\ref{eq:MRmain})  as coefficients to the added mass term, the Stokes drag term and the Basset history integral. In table~\ref{tab:drag} we show how these quantities vary from that of a sphere as the aspect ratio $\k$ is varied. The solution for all time-dependent spatially homogeneous flows can now be easily written for all these geometries as
    \begin{align}
        -\pi \q(0,t)  =& \ \int_{\dD^-} \frac{ik \e^{-k^2 t} \hf(k)} {(\hat{\alpha} - k^2 + ik \hat{\gamma})} \ \d k - \int_{-\infty}^{\infty} \frac{\v_o ik \e^{-k^2 t}} {(\hat{\alpha} - k^2 + ik \hat{\gamma})} \ \d k.
    \end{align}
    The numerical method developed for spatially non-linear time dependent flow fields is usable without any change.
    \begin{table}
        \centering
        \begin{tabular}{l c c}
            Geometry & \ $\cC_A$ & \ $\cC_a$ \\
            \hline
            Sphere & \ $1$ & \ $1$ \\
            Prolate spheroids ($\k>1$) & \ $\displaystyle\frac{2[\k \ln (\k + \sqrt{\k^2 -1}) - \sqrt{\k^2-1}]}{\varkappa^2 \sqrt{\k^2-1} - \k \ln (\k + \sqrt{\k^2 -1})}$ & \ $ \frac{4(\k^2-1)}{3[(2\k^2-1) \ln \big(\k+\sqrt{1-\k^2} \big)/\sqrt{\k^2-1} -\k]} $ \\
            Oblate spheroids ($\k<1$) & \ $\displaystyle\frac{2[\k\cos^{-1} \k - \sqrt{1-\k^2}]}{\k^2\sqrt{1-\k^2} - \k \cos^{-1} \k}$ & \ $\frac{4(1-\k^2)}{3[(1-2\k^2) \cos^{-1} \big( \k/\sqrt{1-\k^2} \big) + \k]}$ \\
            Disk & \ $\displaystyle\bigg(\frac{4}{\pi}\bigg)$ & \ $\displaystyle\bigg(\frac{8}{3\pi} \bigg)$ \\
            \hline
        \end{tabular}
        \caption{Comparison of drag-coefficients~\citep{clift2005} of different particle shapes where $\varkappa$ is the aspect ratio of spheroid, $\cC_A$ is the additional coefficient for added mass term, $\cC_a$ is the additional coefficient for stokes drag, $\cC_H = \cC_a^2$ is in front of Basset history integral.}
        \label{tab:drag}
    \end{table}
    \subsection{Implications to physical scenarios}
    In this paper we explored a range of scenarios, from that of a particle relaxing to zero velocity from a given initial velocity to a particle in a non-linear flow such as the point-vortex. Real flow situations are often describable by a combination of these simple flow situations. Our results on the droplet growth highlight that for raindrop formation rates, the Basset history integral is important.  Our exact solutions for oscillating flow can directly be used to study particle dynamics in synthetic turbulence, defined by a superposition of sinusoidal forcing of different frequencies (see e.g. ~\cite{parmar2018}). As has been shown in~\cite{ravichandran2015} the behaviour of the particle in the presence of a point vortex is directly applicable to 2-D turbulence. The fact that the critical radius for caustic formation shrinks will have implications for droplet clustering.
    \section{Conclusion}
    The Maxey-Riley equation has been used extensively by researchers, but most often by neglecting the history integral, primarily due to the difficulty in handling this term. The non-locality in this term has been the barrier in terms of numerical progress as memory requirements keep increasing with time. This is a major hurdle especially when a large number of particles need to be simulated in high Reynolds number turbulence. Another barrier was that restarting a simulation with history integrals was practically impossible, owing to exponentially large storage requirements. Both these issues have been addressed in this paper by reformulating the Maxey-Riley equation as a boundary condition of the 1-D diffusion equation. Several analytical solutions to the complete Maxey-Riley equation, which can form the basis for gradient expansions in more complicated background flows, have been obtained. Our analysis suggests that the Basset history may not be categorised as an additional drag on a particle. Its effect on particle velocity is seen to cause a more rapid stretched-exponential decay at short times and a power-law at long times. Basset history, we show, has effects on particles settling under gravity, droplet growth rate and migration in a simple shear, in all of which neglecting this term would result in a qualitatively different (and incorrect) prediction. Our numerical scheme is the first spectrally accurate method to our knowledge for the complete Maxey-Riley equation. Its implementation is straightforward, since Chebyshev coefficients can be computed efficiently using the \texttt{FFTW} package. Several open questions, e.g. regarding statistical properties of particles in homogeneous isotropic turbulence can be studied by this approach, and the effect of Basset history in turbulence understood. The larger question of understanding the MR equation from a dynamical systems perspective is as yet unanswered, as is the question of whether attractors are modified by the history integral.\\

We thank Prof. Jacques Magnaudet for important input, and the anonymous referees for comments which improved the paper.

\appendix

    \section{Dirichlet to Neumann map}\label{app:DNmap}
    For the convenience of the reader, we present here the details of the Dirichlet to Neumann map for the diffusion equation. This calculation may also be found in \cite{fokas2008unified}. We begin with the global relation for the 1-D diffusion equation, equation (\ref{eq:gr}), and derive an expression for $\q_x(0,t)$ in terms of $\q(0,t)$. To do so we multiply (\ref{eq:gr}) by $ik\e^{-\omega(k) t}$, for $0<t<T$, and integrate over the contour $\dD^-$ as depicted in figure~\ref{fig:utm_cont}$(b)$. We denote by $\D^-$ ($\D^+$) the regions of the complex$-k$ plane where $\Re(\omega(k)=k^2)<0$ and $\Im(k)<0$ (respectively, $\Im(k)>0$). Note for  $k\in\D^{\pm}$, $\e^{\omega(k) t}$ is bounded and decaying for $t>0$ as $k\to\infty$. We now have
    \begin{equation}
        \int_{\dD^-} [ ik \e^{\omega(k) (T-t)} \hq (k,T) + ik \e^{-\omega(k)t} (\hg_1 + ik \hg_o)]\ \d k = 0.
    \end{equation}
    The first term in the integral affords no contribution. This follows by noting that the integral of this term along $\mathcal{C}_R$, see figure~\ref{fig:utm_cont}$(b)$, vanishes as $R \rightarrow \infty$ for $(T-t)>0$. Moreover since there are no poles in $\D^-$, the integral of the first term along the contour $\dD^-$ is also zero. Next we employ the definition of $\hg_0$ and $\hg_1$ to obtain
    \begin{align}
        \int_{\dD^-} \bigg[ ik \int_o^T \e^{\omega(k) (s-t)} \q_x (0,s) \ \d s - k^2 \int_o^T \e^{\omega(k) (s-t) } \q (0,s) \ \d s  \bigg] \ \d k =& \ 0.
    \end{align}
    The integrand of the contour integral in the above expression consists of the sum of two terms. For the first term we substitute $k^2=il,\ l\in\mathbb{R}$ to obtain
    \begin{align}
        -\fot \int_0^T \int_{-\infty}^{\infty} \e^{-il (t-s)} \q_x (0,s) \ \d l \ \d s - \int_{\dD^-} k^2 \bigg( \int_0^T \e^{k^2 (s-t)} \q (0,s) \ \d s \bigg) \ \d k =& \ 0.
    \end{align}
    The Fourier inversion theorem then allows us to replace the first term with the Neumann condition $\q_x(0,t)$. The second term on the left-hand side may be simplified as follows. First we integrate by parts once and then deform the integral from $\dD^-$ to the real line
    \begin{align}
        &-\pi \q_x (0,t) - \int_{\dD^-} k^2 \bigg[ \bigg( \frac{\e^{k^2 (T-t)} \q (0,T) - \e^{-k^2 t} \q (0,0)}{k^2} \bigg) - \int_0^T \e^{k^2 (s-t)} \dot{\q} (0,s) \ \d s \bigg] \ \d k = \ 0,\\
        \Rightarrow 
        &-\pi \q_x (0,t) - \int_{-\infty}^{\infty} \e^{-k^2 t} \q (0,0)\ \d k \nonumber \\  &\quad \quad- \int_{\dD^-} \bigg[ \e^{k^2 (T-t)} \q (0,T) -  \bigg(\int_0^t + \int_t^T \bigg)  \e^{k^2 (s-t)} \dot{\q} (0,s) \ \d s \bigg] \ \d k = \ 0,	\label{eq:dnderiv}\\
        \Rightarrow &-\pi \q_x (0,t) - \sqrt{\frac{\pi}{t}} \q (0,0) - \int_{-\infty}^{\infty} \int_0^t \e^{k^2 (s-t)} \dot{\q} (0,s) \ \d s \ \d k = \ 0. \label{eq:dnderiv1}
    \end{align}
    The last term of (\ref{eq:dnderiv}) consists of an integral over $s\in (t,T]$. This term vanishes since for $(s-t)>0$, $\e^{k^2(s-t)}$ is analytic, bounded and decays at infinity for $k\in \D^-$. An appeal to Jordan's lemma assures us that the contribution from this term is zero. A similar argument implies the term with $\q(0,T)$ also vanishes. Switching the order of integrations for the integral term in  (\ref{eq:dnderiv1}) leads to the final expression for the Dirichlet to Neumann map
    \begin{align}	
        \q_x (0,t) =& - \sqrt{\frac{1}{\pi t}} \q (0,0) - \frac{1}{\sqrt{\pi}} \int_0^t \frac{\dot{\q} (0,s)}{\sqrt{t-s}} \ \d s.
    \end{align}
    \section{History evolution}\label{ssec:histevol}
    Multiplying the boundary condition (\ref{eq:BCmod}) by $\e^{k^2t}$ and integrating over $t\in[t_o,t_o+\Dt]$ leads to 
    \begin{align} \nonumber
        \e^{k^2 (t_o+\Delta T) } \q(0,t_o+\Delta t) -& \e^{k^2 t_o} \q(0,t_o) + (\alpha - k^2)\hg_o(k^2,t_o,t_o+\Delta t)\\  &\quad\quad - \gamma \hg_1(k^2,t_o,t_o+\Delta t) = \ \hf(k^2,t_o,t_o+\Delta t).\label{eq:BCmod_transform}
    \end{align}
    We may now eliminate $\hg_1(k^2,t_o,t_o+\Delta t)$ in (\ref{eq:GRuseful}) using the above expression. Multiplying the resulting expression by $k \e^{-k^2 t}/(\alpha - k^2 + ik\gamma)$ and integrating over $\dD^-$ results in the following relation for the Dirichlet condition $\q(0,t)$
    \begin{align}\nonumber
        i\pi \q(0,t) = \ \q(0,t_o) \int_{\partial \D^-} & \frac{k \e^{-k^2(t-t_o)}}{(\alpha - k^2 + ik\gamma)} \ \d k  +  \int_{\dD^-} \frac{k \e^{-k^2 t} \hf(k^2,t_o,t_o+\Delta t)}{(\alpha - k^2 + ik\gamma)} \ \d k \\
        &  + \int_{\dD^-} \frac{k \e^{-k^2(t-t_o)}\hq(k,t_o)}{(\alpha - k^2 + ik\gamma)} \ \d k,\quad \ t_o \leq t \leq t_o+\Delta t. \label{eq:appH1}
    \end{align}
    To obtain the above expression we appeal to Jordan's lemma several times in order to eliminate integrals that vanish. Equation (\ref{eq:appH1}) may be deformed back to the real line to obtain equation (\ref{eq:fin1}).  

    Equation (\ref{eq:appH1}) expresses $\q(0,t)$ in terms of an initial condition $\q(0,t_o)$, itself (via $\hf$) and $\hq(k,t_o)$. We now derive an expression to relate $\hq(k,t_o)$ and $\hq(k,t_o+\Delta t)$ in order to obtain a rule to update the $\hq$. Once again, eliminating $\hg_1$ from the global relation (\ref{eq:GRuseful}) and (\ref{eq:BCmod_transform}) we obtain after a bit of algebra 
    \begin{align}
        \e^{k^2(t_o+\Delta t)} \H(k,t_o+\Delta t) +  \hg_o(k^2,t_o,t_o+\Delta t) =& \ \e^{k^2 t_o} \H(k,t_o) + \frac{\hf(k^2,t_o,t_o+\Delta t)}{(\alpha - k^2 + ik \ga)}, \label{eq:RGhist}
    \end{align}
    where
    $$
    \H(k,t_o) = \bigg\{ \frac{\q(0,t_o)+ \gamma \hq(k,t_o)}{(\alpha - k^2 + ik \ga)} \bigg\},
    $$
    which at the initial time instant is $\H(k,0) = \q(0,0)/(\alpha - k^2 + ik \ga)$. A similar relation exists between $\H(k,t_o-\Dt)$ and $\H(k,t_o)$ (obtained by considering (\ref{eq:BCmod_transform}) and (\ref{eq:GRuseful}) for the time interval $[t_o-\Dt,t_o]$ ) which is precisely (\ref{eq:fin2}).

    \section{Evaluating individual terms in equation (\ref{eq:fin1})} \label{app:Ntech}
    \begin{figure}
        \centering
        \includegraphics{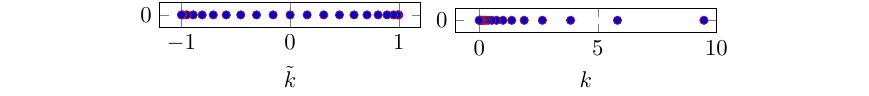}
        \caption{Map of variable $k \rightarrow \tk$, where $\tk_l$ are Gauss-Lobatto grid-points, the discrete version of $\tk$, which is used to evaluate integral in equation (\ref{eq:stand}).} \label{fig:grid}
    \end{figure}
    In order to reach equation (\ref{eq:march}), there are four steps in the process of building $\His(t_j)$ and $\F(t_j)$. We start with the first step which is to evaluate indefinite integrals. In order to evaluate integrals of the following type numerically
    \begin{align}
        \mathscr{P}(t) :=& \ \int_{-\infty}^{\infty} \mathscr{B}(k) \e^{-k^2t} \ \d k = \ \int_{0}^{\infty} \Im \big\{ \mathscr{B}(k) \big\} \e^{-k^2t} \ \d k. \label{eq:stand}
    \end{align}
    Integrals of this form are prevalent in both equation (\ref{eq:fin1}-\ref{eq:fin2}) for smooth functions $\mathscr{B}(k)$ which are well behaved in the entire interval $k \in (-\infty,\infty)$. We first map the variable $k \rightarrow \tk$ using the transformation:
    $
    k = {(1+\tk)}/{(1-\tk)}, \ \tk \in [-1,1).
    $
    Using this map, we can expand the function $\Im\big\{ \mathscr{B}(k)\big\}$ using Chebyshev basis as: $\sum_{n=0}^N c_n T_n(\tk)$. Using Fast Fourier Transform (\texttt{FFT}) packages, we compute $c_n$ and evaluate the integration $\mathscr{P}(t)$. Thus evaluating this integral just requires one forward Chebyshev Transform (\texttt{CT}). We show in figure~\ref{fig:grid} the map between Gauss-Lobatto grid points $\tk_l \in [-1,1)$ and $k_l \in [0,\infty)$. Using this method we can evaluate the entire expression $\His(t_j)$. 

    We move to the second step in evaluating the equation (\ref{eq:fin1}) which is the term $\F(t_j)$. For a given guess $\q(0,t_j)$, which we take as $-2/\pi \His(t_j)$, we can evaluate $\f(\q(0,t_j),t_j)$. The next step is then to compute from this given $\q(0,t_j)$ the integral $\F(t)$:
    \begin{align}
        \F(t) =& \ \int_{0}^\infty \Im \bigg\{ \frac{k \e^{-k^2t}\hf(k^2,t_o,t)}{(\alpha - k^2 + ik\ga)} \bigg\} \ \d k ,\quad t \in [t_o, t_o+\Delta t], \\
        =& \ \int_{t_o}^t \f(\q(0,s)) \int_{0}^\infty \Im \bigg\{\frac{k \e^{-k^2(t-s)}}{(\alpha - k^2 + ik\ga)} \bigg\} \ \d k \ \d s.
    \end{align}
    Substituting $(t-s) = m$, we get:
    \begin{align}
        \F(t) =& \ \int_0^{t-t_o} \f(\q(0,t-m)) \underbrace{\int_{0}^\infty \Im \bigg\{\frac{k \e^{-k^2(t-s)}}{(\alpha - k^2 + ik\ga)} \bigg\} \ \d k}_{\mathscr{L}(m)} \ \d m.
    \end{align}
    We can now expand $\f(\q(0,t-m)) = \sum_{n=0}^N c_n T_n ((t-m)-(t_o+\Delta t/2)/(\Delta t/2))$ with $c_n$ being function of $\q(0,t_j)$. We choose $t-t_o = \tt$, which eliminates the $t_o$ dependence from the expansion and this lets us write
    \begin{align}
        \F(t_j) =& \ \sum_{l=0}^N c_n \int_0^{\tt_j} T_n \bigg(\frac{(\tt_j-m)-\Delta t/2}{\Delta t/2} \bigg) \Lscr(m) \ \d m, \ \tt_j \in [0,\Delta t], \\
        =& \ \sum_{l=0}^N c_n \M_{nj}.
    \end{align}
    We can precompute $\M_{nj}$, which is a $N\times N$ matrix, and evaluating $\F(t_j)$ again requires only one $\texttt{CT}$.	The next term in equation (\ref{eq:fin2}) needs a small trick to simplify and is given by
    \begin{align}
        \I_1 =& \ \int_{t_o - \Dt}^{t_o} \e^{-k^2(t_o-s)} \q(0,s) \ \d s, \\
        =& \ \Dt \int_{0}^{1} \e^{-k^2\tau} \q(0,t_o-\Dt \tau) \ \d \tau, \ \text{where} \ s=t_o-\Dt \tau,\\
        =& \ \frac{\Dt}{k^2} \int_{0}^{k^2} \e^{-\lambda} \q(0,t_o-\Dt \frac{\lambda}{k^2}) \ \d \lambda, \ \text{with} \ \tau=\frac{\lambda}{k^2}.
    \end{align}
    We immediately see from this expression that for large $k$, the integral goes to: $\q(0,t_o)\Dt/k^2 + \O(1/k^2)$ implying that this part can be separately evaluated by Taylor expansion around $t_o$. This gives rise to
    \begin{align}
        \I_1 =& \begin{cases}
            \int_{t_o - \Dt}^{t_o} \q(0,s) \ \d s, & \text{if } k = 0 \\[2em]
            \frac{\Dt}{k^2} \bigg[ \q(0,t_o) (1-\e^{-k^2}) + \int_{0}^{k^2} \e^{-\lambda} \bigg( \q(0,t_o-\Dt \frac{\lambda}{k^2}) - \q(0,t_o) \bigg) \ \d \lambda \bigg] , & \text{if } k \ne 0
        \end{cases}\,
    \end{align}
    The second term in the above equation is evaluated by expanding $\q(0,t)$ using values at $t=t_j$ when $ t_o \leq t_j \leq (t_o + \Dt)$ as $\sum_{n=0}^N a_n T_n(\tt)$ where $\tt =(2t-(2t_o+\Dt))/\Dt$ is the map from $t \rightarrow \tt$. Given this polynomial approximation, we can write the integral as a precomputed matrix $\mathbb{D}_{nm}$ times the $\texttt{CT}$ vector of $\q(0,t_j)$, $a_n$. The subsequent integral $\I_2$ follows the same argument and we can use the exact same matrix $\mathbb{D}_{nm}$ to compute it. Thus to compute $\mathscr{T}(t_j)$, all that we need is precomputed matrices $\mathbb{D}_{nm}$ and $\M_{nj}$ beyond which it is computing two $\texttt{CT}$s for each time, $t_o \leq t_j \leq (t_o + \Dt)$.

    \bibliographystyle{jfm}
    \bibliography{biblio}

\begin{thebibliography}{32}
\expandafter\ifx\csname natexlab\endcsname\relax\def\natexlab#1{#1}\fi
\def\au#1{#1} \def\ed#1{#1} \def\yr#1{#1}\def\at#1{#1}\def\jt#1{\textit{#1}}
  \def\bt#1{#1}\def\bvol#1{\textbf{#1}} \def\vol#1{#1} \def\pg#1{#1}
  \def\publ#1{#1}\def\arxiv#1{#1}\def\org#1{#1}\def\st#1{\textit{#1}}

\bibitem[Ablowitz \& Fokas(2003)]{ablowitz2003complex}
{\sc \au{Ablowitz, Mark~J} \& \au{Fokas, Athanassios~S}} \yr{2003} {\em Complex
  variables: introduction and applications\/}.  \publ{Cambridge University
  Press}.

\bibitem[Ardekani {\em et~al.\/}(2016)Ardekani, Sardina, Brandt, Karp-Boss,
  Bearon \& Variano]{ardekani2016sedimentation}
{\sc \au{Ardekani, M~Niazi}, \au{Sardina, G}, \au{Brandt, L}, \au{Karp-Boss,
  L}, \au{Bearon, RN} \& \au{Variano, EA}} \yr{2016}  \at{Sedimentation of
  elongated non-motile prolate spheroids in homogenous isotropic turbulence}.
  \jt{arXiv preprint arXiv:1611.10284} .

\bibitem[Basset(1888)]{basset1888treatise}
{\sc \au{Basset, Alfred~Barnard}} \yr{1888} {\em A treatise on hydrodynamics:
  with numerous examples\/}, ,  \vol{vol.~2}.  \publ{Deighton, Bell and
  Company}.

\bibitem[Clift {\em et~al.\/}(2005)Clift, Grace \& Weber]{clift2005}
{\sc \au{Clift, Roland}, \au{Grace, John~R} \& \au{Weber, Martin~E}} \yr{2005}
  {\em Bubbles, drops, and particles\/}.  \publ{Courier Corporation}.

\bibitem[Daitche(2013)]{daitche2013}
{\sc \au{Daitche, Anton}} \yr{2013}  \at{Advection of inertial particles in the
  presence of the history force: Higher order numerical schemes}.  \jt{Journal
  of Computational Physics}  \bvol{254},  \pg{93--106}.

\bibitem[Daitche(2015)]{daitche2015role}
{\sc \au{Daitche, Anton}} \yr{2015}  \at{On the role of the history force for
  inertial particles in turbulence}.  \jt{Journal of Fluid Mechanics}
  \bvol{782},  \pg{567--593}.

\bibitem[Daitche \& T{\'e}l(2011)]{daitche2011memory}
{\sc \au{Daitche, Anton} \& \au{T{\'e}l, Tam{\'a}s}} \yr{2011}  \at{Memory
  effects are relevant for chaotic advection of inertial particles}.
  \jt{Physical review letters}  \bvol{107}~(24),  \pg{244501}.

\bibitem[Daitche \& T{\'e}l(2014)]{daitche2014memory}
{\sc \au{Daitche, Anton} \& \au{T{\'e}l, Tam{\'a}s}} \yr{2014}  \at{Memory
  effects in chaotic advection of inertial particles}.  \jt{New Journal of
  Physics}  \bvol{16}~(7),  \pg{073008}.

\bibitem[Deconinck {\em et~al.\/}(2017)Deconinck, Guo, Shlizerman \&
  Vasan]{deconinck2017}
{\sc \au{Deconinck, Bernard}, \au{Guo, Qi}, \au{Shlizerman, Eli} \& \au{Vasan,
  Vishal}} \yr{2017}  \at{Fokas's unified transform method for linear systems}.
   \jt{Quarterly of Applied Mathematics} .

\bibitem[Deconinck {\em et~al.\/}(2014)Deconinck, Trogdon \&
  Vasan]{deconinck2014}
{\sc \au{Deconinck, Bernard}, \au{Trogdon, Thomas} \& \au{Vasan, Vishal}}
  \yr{2014}  \at{The method of fokas for solving linear partial differential
  equations}.  \jt{SIAM Review}  \bvol{56}~(1),  \pg{159--186}.

\bibitem[Deepu {\em et~al.\/}(2017)Deepu, Ravichandran \&
  Govindarajan]{deepu2017caustics}
{\sc \au{Deepu, P}, \au{Ravichandran, S} \& \au{Govindarajan, Rama}} \yr{2017}
  \at{Caustics-induced coalescence of small droplets near a vortex}.
  \jt{Physical Review Fluids}  \bvol{2}~(2),  \pg{024305}.

\bibitem[Elghannay \& Tafti(2016)]{elghannay2016development}
{\sc \au{Elghannay, Husam~A} \& \au{Tafti, Danesh~K}} \yr{2016}
  \at{Development and validation of a reduced order history force model}.
  \jt{International Journal of Multiphase Flow}  \bvol{85},  \pg{284--297}.

\bibitem[Falkovich {\em et~al.\/}(2002)Falkovich, Fouxon \&
  Stepanov]{falkovich2002}
{\sc \au{Falkovich, G}, \au{Fouxon, A} \& \au{Stepanov, MG}} \yr{2002}
  \at{Acceleration of rain initiation by cloud turbulence}.  \jt{Nature}
  \bvol{419}~(6903),  \pg{151}.

\bibitem[Farazmand \& Haller(2015)]{farazmand2015}
{\sc \au{Farazmand, Mohammad} \& \au{Haller, George}} \yr{2015}  \at{The
  maxey--riley equation: Existence, uniqueness and regularity of solutions}.
  \jt{Nonlinear Analysis: Real World Applications}  \bvol{22},  \pg{98--106}.

\bibitem[Fokas(2008)]{fokas2008unified}
{\sc \au{Fokas, A.~S.}} \yr{2008} {\em {A unified approach to boundary value
  problems}\/}.  \publ{Philadelphia, PA: CBMS-NSF regional conference series in
  applied mathematics, Society for Industrial and Applied Mathematics (SIAM)}.

\bibitem[Govindarajan \& Ravichandran(2017)]{croor2017}
{\sc \au{Govindarajan, Rama} \& \au{Ravichandran, S.}} \yr{2017}  \at{Cloud
  microatlas}.  \jt{Resonance}  \bvol{22}~(3),  \pg{269--277}.

\bibitem[Guseva {\em et~al.\/}(2016)Guseva, Daitche, Feudel \&
  T{\'e}l]{guseva2016history}
{\sc \au{Guseva, Ksenia}, \au{Daitche, Anton}, \au{Feudel, Ulrike} \&
  \au{T{\'e}l, Tam{\'a}s}} \yr{2016}  \at{History effects in the sedimentation
  of light aerosols in turbulence: The case of marine snow}.  \jt{Physical
  Review Fluids}  \bvol{1}~(7),  \pg{074203}.

\bibitem[Guseva {\em et~al.\/}(2017)Guseva, Daitche \&
  T{\'e}l]{guseva2017snapshot}
{\sc \au{Guseva, Ksenia}, \au{Daitche, Anton} \& \au{T{\'e}l, Tam{\'a}s}}
  \yr{2017}  \at{A snapshot attractor view of the advection of inertial
  particles in the presence of history force}.  \jt{The European Physical
  Journal Special Topics}  \bvol{226}~(9),  \pg{2069--2078}.

\bibitem[Klinkenberg {\em et~al.\/}(2014)Klinkenberg, de~Lange \&
  Brandt]{klinkenberg2014}
{\sc \au{Klinkenberg, Joy}, \au{de~Lange, HC} \& \au{Brandt, Luca}} \yr{2014}
  \at{Linear stability of particle laden flows: the influence of added mass,
  fluid acceleration and basset history force}.  \jt{Meccanica}  \bvol{49}~(4),
   \pg{811--827}.

\bibitem[Langlois {\em et~al.\/}(2015)Langlois, Farazmand \&
  Haller]{langlois2015}
{\sc \au{Langlois, Gabriel~Provencher}, \au{Farazmand, Mohammad} \& \au{Haller,
  George}} \yr{2015}  \at{Asymptotic dynamics of inertial particles with
  memory}.  \jt{Journal of nonlinear science}  \bvol{25}~(6),  \pg{1225--1255}.

\bibitem[Lovalenti \& Brady(1993{\natexlab{{\em a\/}}})]{lovalenti1993b}
{\sc \au{Lovalenti, Phillip~M} \& \au{Brady, John~F}} \yr{1993{\natexlab{{\em
  a\/}}}}  \at{The force on a sphere in a uniform flow with small-amplitude
  oscillations at finite reynolds number}.  \jt{Journal of Fluid Mechanics}
  \bvol{256},  \pg{607--614}.

\bibitem[Lovalenti \& Brady(1993{\natexlab{{\em b\/}}})]{lovalenti1993}
{\sc \au{Lovalenti, Phillip~M} \& \au{Brady, John~F}} \yr{1993{\natexlab{{\em
  b\/}}}}  \at{The hydrodynamic force on a rigid particle undergoing arbitrary
  time-dependent motion at small reynolds number}.  \jt{Journal of Fluid
  Mechanics}  \bvol{256},  \pg{561--605}.

\bibitem[Maxey \& Riley(1983)]{maxey1983}
{\sc \au{Maxey, Martin~R} \& \au{Riley, James~J}} \yr{1983}  \at{Equation of
  motion for a small rigid sphere in a nonuniform flow}.  \jt{The Physics of
  Fluids}  \bvol{26}~(4),  \pg{883--889}.

\bibitem[Mei \& Adrian(1992)]{mei1992}
{\sc \au{Mei, Renwei} \& \au{Adrian, Ronald~J}} \yr{1992}  \at{Flow past a
  sphere with an oscillation in the free-stream velocity and unsteady drag at
  finite reynolds number}.  \jt{Journal of Fluid Mechanics}  \bvol{237},
  \pg{323--341}.

\bibitem[Miller(2006)]{miller2006}
{\sc \au{Miller, Peter~David}} \yr{2006} {\em Applied asymptotic analysis\/}, ,
   \vol{vol.~75}.  \publ{American Mathematical Soc.}

\bibitem[Olivieri {\em et~al.\/}(2014)Olivieri, Picano, Sardina, Iudicone \&
  Brandt]{olivieri2014effect}
{\sc \au{Olivieri, Stefano}, \au{Picano, Francesco}, \au{Sardina, Gaetano},
  \au{Iudicone, Daniele} \& \au{Brandt, Luca}} \yr{2014}  \at{The effect of the
  basset history force on particle clustering in homogeneous and isotropic
  turbulence}.  \jt{Physics of fluids}  \bvol{26}~(4),  \pg{041704}.

\bibitem[Parmar {\em et~al.\/}(2018)Parmar, Annamalai, Balachandar \&
  Prosperetti]{parmar2018}
{\sc \au{Parmar, M}, \au{Annamalai, S}, \au{Balachandar, S} \& \au{Prosperetti,
  A}} \yr{2018}  \at{Differential formulation of the viscous history force on a
  particle for efficient and accurate computation}.  \jt{Journal of Fluid
  Mechanics}  \bvol{844},  \pg{970--993}.

\bibitem[Ravichandran \& Govindarajan(2015)]{ravichandran2015}
{\sc \au{Ravichandran, S} \& \au{Govindarajan, Rama}} \yr{2015}  \at{Caustics
  and clustering in the vicinity of a vortex}.  \jt{Physics of Fluids}
  \bvol{27}~(3),  \pg{033305}.

\bibitem[Ravichandran \& Govindarajan(2017)]{ravichandran2017}
{\sc \au{Ravichandran, S} \& \au{Govindarajan, Rama}} \yr{2017}
  \at{Vortex-dipole collapse induced by droplet inertia and phase change}.
  \jt{Journal of Fluid Mechanics}  \bvol{832},  \pg{745--776}.

\bibitem[Toschi \& Bodenschatz(2009)]{toschi2009}
{\sc \au{Toschi, Federico} \& \au{Bodenschatz, Eberhard}} \yr{2009}
  \at{Lagrangian properties of particles in turbulence}.  \jt{Annual review of
  fluid mechanics}  \bvol{41},  \pg{375--404}.

\bibitem[Van~Hinsberg {\em et~al.\/}(2011)Van~Hinsberg, ten Thije~Boonkkamp \&
  Clercx]{van2011}
{\sc \au{Van~Hinsberg, MAT}, \au{ten Thije~Boonkkamp, JHM} \& \au{Clercx,
  Hans~JH}} \yr{2011}  \at{An efficient, second order method for the
  approximation of the basset history force}.  \jt{Journal of Computational
  Physics}  \bvol{230}~(4),  \pg{1465--1478}.

\bibitem[Vasan {\em et~al.\/}(2018)Vasan, Ganga~Prasath \&
  Govindarajan]{vishal2018}
{\sc \au{Vasan, Vishal}, \au{Ganga~Prasath, S} \& \au{Govindarajan, Rama}}
  \yr{2018}  \at{Wellposedness of fractional differential equations with
  applications to the maxey-riley equation}.  \jt{In preparation} .

\end{thebibliography}


\begin{thebibliography}{14}
\expandafter\ifx\csname natexlab\endcsname\relax\def\natexlab#1{#1}\fi
\def\au#1{#1} \def\ed#1{#1} \def\yr#1{#1}\def\at#1{#1}\def\jt#1{\textit{#1}}
  \def\bt#1{#1}\def\bvol#1{\textbf{#1}} \def\vol#1{#1} \def\pg#1{#1}
  \def\publ#1{#1}\def\arxiv#1{#1}\def\org#1{#1}\def\st#1{\textit{#1}}

\bibitem[Batchelor(1971)]{Batchelor59}
{\sc \au{Batchelor, G.~K.}} \yr{1971}  \at{Small-scale variation of convected
  quantities like temperature in turbulent fluid. part 1. general discussion
  and the case of small conductivity.}  \jt{J.~Fluid Mech.}  \bvol{5},
  \pg{113--133}.

\bibitem[Brownell \& Su(2004)]{Brownell04}
{\sc \au{Brownell, C.~J.} \& \au{Su, L.~K.}} \yr{2004}  \at{Planar measurements
  of differential diffusion in turbulent jets}.  \jt{AIAA Paper 2004-2335} .

\bibitem[Brownell \& Su(2007)]{Brownell07}
{\sc \au{Brownell, C.~J.} \& \au{Su, L.~K.}} \yr{2007}  \at{Scale relations and
  spatial spectra in a differentially diffusing jet}.  \jt{AIAA Paper
  2007-1314} .

\bibitem[Dennis(1985)]{Dennis85}
{\sc \au{Dennis, S. C.~R.}} \yr{1985}  \at{{Compact explicit finite difference
  approximations to the Navier--Stokes equation}}.  \bt{In {\em Ninth Intl
  Conf. on Numerical Methods in Fluid Dynamics\/} (ed. \ed{Soubbaramayer \&
  J.~P. Boujot})},  \st{Lecture Notes in Physics},  \vol{vol. 218},  \pg{pp.
  23--51}.  \publ{Springer}.

\bibitem[Hwang \& Tuck(1970)]{Hwang70}
{\sc \au{Hwang, L.-S.} \& \au{Tuck, E.~O.}} \yr{1970}  \at{On the oscillations
  of harbours of arbitrary shape}.  \jt{J.~Fluid Mech.}  \bvol{42},
  \pg{447--464}.

\bibitem[Koch(1983)]{Koch83}
{\sc \au{Koch, W.}} \yr{1983}  \at{Resonant acoustic frequencies of flat plate
  cascades}.  \jt{J.~Sound Vib.}  \bvol{88},  \pg{233--242}.

\bibitem[Lee(1971)]{Lee71}
{\sc \au{Lee, J.-J.}} \yr{1971}  \at{Wave-induced oscillations in harbours of
  arbitrary geometry}.  \jt{J.~Fluid Mech.}  \bvol{45},  \pg{375--394}.

\bibitem[Linton \& Evans(1992)]{Linton92}
{\sc \au{Linton, C.~M.} \& \au{Evans, D.~V.}} \yr{1992}  \at{The radiation and
  scattering of surface waves by a vertical circular cylinder in a channel}.
  \jt{Phil.\ Trans.\ R. Soc.\ Lond.}  \bvol{338},  \pg{325--357}.

\bibitem[Martin(1980)]{Martin80}
{\sc \au{Martin, P.~A.}} \yr{1980}  \at{On the null-field equations for the
  exterior problems of acoustics}.  \jt{Q.~J. Mech.\ Appl.\ Maths}  \bvol{33},
  \pg{385--396}.

\bibitem[Miller(1991)]{Miller91}
{\sc \au{Miller, P.~L.}} \yr{1991}  \at{Mixing in high schmidt number turbulent
  jets}. PhD thesis, California Institute of Technology.

\bibitem[Rogallo(1981)]{Rogallo81}
{\sc \au{Rogallo, R.~S.}} \yr{1981}  \bt{Numerical experiments in homogeneous
  turbulence}. {\em Tech. Rep.\/} 81835.  \org{NASA Tech.\ Mem.}

\bibitem[Ursell(1950)]{Ursell50}
{\sc \au{Ursell, F.}} \yr{1950}  \at{Surface waves on deep water in the
  presence of a submerged cylinder i}.  \jt{Proc.\ Camb.\ Phil.\ Soc.}
  \bvol{46},  \pg{141--152}.

\bibitem[{van Wijngaarden}(1968)]{Wijngaarden68}
{\sc \au{{van Wijngaarden}, L.}} \yr{1968}  \at{On the oscillations near and at
  resonance in open pipes}.  \jt{J.~Engng Maths}  \bvol{2},  \pg{225--240}.

\bibitem[Worster(1992)]{Worster92}
{\sc \au{Worster, M.~G.}} \yr{1992}  \at{{The dynamics of mushy layers}}.
  \bt{In {\em In Interactive dynamics of convection and solidification\/} (ed.
  \ed{S.~H. Davis, H.~E. Huppert, W.~Muller \& M.~G. Worster})},  \pg{pp.
  113--138}.  \publ{Kluwer}.

\end{thebibliography}
\end{document}